\documentclass[prd,amsfonts,aps,nofootinbib,notitlepage,11pt,superscriptaddress]{revtex4-1}

\pdfoutput=1
\usepackage{amsmath,amssymb}
\usepackage{graphicx}
\usepackage[utf8]{inputenc}
\usepackage{cancel}
\usepackage[colorlinks=true]{hyperref}
\usepackage{color}
\usepackage{array,multirow}
\usepackage{subcaption}
\usepackage{float}
\usepackage{comment}

\newcommand{\UPTC}{Escuela de Física, Universidad Pedagógica y Tecnológica de Colombia,\\
Avenida Central del Norte \# 39-115, Tunja, Colombia}
\newcommand{\UdeA}{Instituto de Física, Universidad de Antioquia,\\Calle 70 \# 52-21, Apartado Aéreo 1226, Medellín, Colombia}

\begin{document}
\title{Fermion and scalar two-component dark matter from a $Z_4$ symmetry}

\author{Carlos E. Yaguna}
\affiliation{\UPTC}
\author{\'Oscar Zapata}
\affiliation{\UdeA}

\begin{abstract}
We revisit a  two-component dark matter model in which the dark matter particles are a singlet fermion ($\psi$) and a singlet scalar ($S$), both stabilized by a single $Z_4$ symmetry. The model, which was proposed by Y. Cai and A. Spray, is remarkably simple, with its phenomenology determined by just five parameters: the two dark matter masses and three dimensionless couplings. In fact, $S$ interacts with the Standard Model particles via the usual Higgs-portal, whereas $\psi$ only interacts directly with $S$, via the Yukawa terms $\overline{\psi^c}(y_s+y_p\gamma^5)\psi\,S$. We consider the two possible mass hierarchies among the dark matter particles, $M_S<M_\psi$ and $M_\psi<M_S$, and numerically investigate the consistency of the model with current bounds. The main novelties of our analysis are the inclusion of the $y_p$ coupling, the update of the direct detection limits, and a more detailed characterization of the viable parameter space. For dark matter masses below $1.3$ TeV or so, we find that the model not only is compatible  with all known constraints, but that it also gives rise to observable signals in future dark matter experiments. Our results show that both dark matter particles may be observed in  direct detection experiments and that the most relevant indirect detection channel is due to the annihilation of $\psi$. We also argue that this setup can be extended to other $Z_N$ symmetries and additional dark matter particles. 
  \end{abstract}

\maketitle
\section{Introduction}
Determining the nature of the dark matter --that exotic form of matter that accounts for about $25\%$ of the energy density of the Universe \cite{Aghanim:2018eyx}-- is one of the most important open problems in fundamental physics today. A common approach is to assume that the dark matter is explained by \emph{one} elementary particle, which, being neutral and stable, is not part of  the Standard Model (SM)~\cite{Jungman:1995df,Bertone:2004pz}. Throughout the years, many different models have been proposed along these lines~\cite{Arcadi:2017kky,Bernal:2017kxu}.

A simple alternative to this approach is that of  multi-component dark matter scenarios~~\cite{Boehm:2003ha,Ma:2006uv,Cao:2007fy,Hur:2007ur,Lee:2008pc,Zurek:2008qg,Barger:2008jx,Profumo:2009tb,Batell:2010bp,Belanger:2011ww,Baer:2011hx,Liu:2011aa,Ivanov:2012hc,Belanger:2012vp,Modak:2013jya,Belanger:2014vza,Esch:2014jpa,Belanger:2014bga,Cai:2015zza,Biswas:2015sva,Arcadi:2016kmk,Bhattacharya:2016ysw,Bhattacharya:2017fid,Pandey:2017quk,Ahmed:2017dbb,Bhattacharya:2018cgx,YaserAyazi:2018lrv,Bernal:2018aon,Poulin:2018kap,Carvajal:2018ohk,Borah:2019aeq,Nanda:2019nqy,Yaguna:2019cvp,Betancur:2020fdl,Hernandez-Sanchez:2020aop,Belanger:2020hyh,Choi:2021yps,Belanger:2021lwd,Yaguna:2021vhb,DiazSaez:2021pmg,Carvajal:2021fxu,Mohamadnejad:2021tke}, in which  the dark matter consists of several particles, each contributing just a fraction of the observed dark matter density. These scenarios are consistent with current observations and often feature distinctive experimental signatures that allow to differentiate them from the standard setup.   Recently, it was pointed out~\cite{Yaguna:2019cvp} that multi-component \emph{scalar} dark matter models based on  a single $Z_N$ ($N\geq 4$) stabilizing symmetry are well-motivated and offer an interesting phenomenology~\cite{Batell:2010bp,Belanger:2012vp,Belanger:2014bga}. Two-component dark matter scenarios  of this type were studied in Refs.~\cite{Belanger:2020hyh} and \cite{Yaguna:2021vhb}. Here, we widen such discussion to models where the dark matter consists of a scalar and a fermion.

Specifically, we revisit the model proposed in \cite{Cai:2015zza}, which is based on a $Z_4$ symmetry and extends the SM particle content with a Dirac fermion ($\psi$) and a real scalar ($S$), both singlets under the gauge group but charged under the $Z_4$. This model turns out to be remarkably simple, with just five parameters dictating its phenomenology --the two dark matter masses and three couplings. In this paper, we expand and update the analysis of this model in multiple ways. Among others, we include, for the first time, the pseudoscalar coupling $y_p$, which opens up new regions of parameter space; we take into account the most recent limits from dark matter direct detection experiments, which exclude a significant fraction of previously considered viable models; we obtain the viable regions, and characterize them in detail by projecting them onto different planes; we study the most relevant experimental signatures in direct and indirect dark matter experiments; and we show how this model can be straightforwardly extended  to other $Z_N$ symmetries and additional dark matter particles.    
We find that the model is viable over a wide range of masses and that it is experimentally very promising. A novel and crucial result  of our analysis  is that \emph{both} dark matter particles could be observed in current and planned direct detection experiments.

This $Z_4$ model has several advantages: it is likely the simplest two-component dark matter model that can be conceived; it can be seen as a minimal extension of the well-known scalar singlet model~\cite{Silveira:1985rk,McDonald:1993ex,Burgess:2000yq}, with the benefit of remaining viable for dark matter masses below $1$ TeV or so ~\cite{Cline:2013gha,Athron:2018ipf}; it leads to  observable signatures that allow to differentiate it from the more conventional models; and, as will be shown, it belongs to a family of multi-component  models featuring scalar and fermionic dark matter particles that are stabilized by a single $Z_N$ symmetry. 

The rest of the paper is organized as follows. The model is presented in the next section. In section \ref{sec:DMpheno} the dark matter phenomenology is discussed in detail, including the new processes that contribute to the relic densities  and the Boltzmann equations that determine them.  Our main results are presented in sections \ref{sec:pheno} and \ref{sec:pheno2}, where a random scan is used to identify the viable regions of this model for each of the two mass regimes. The direct and indirect detection prospects are also analyzed there. In section \ref{sec:discussion} we briefly examine possible extensions of this model to other $Z_N$ symmetries and to additional dark matter particles. Finally, we draw our conclusion in section \ref{sec:conclusions}. 

\section{The model}\label{sec:model}
Let us consider an extension of the SM by  a real scalar singlet $S$ and a Dirac fermion singlet $\psi$, both charged under a new $Z_4$ symmetry. $S$ and $\psi$ are assumed to transform respectively as $S\to -S$ and $\psi\to i\psi$, whereas the SM fields are singlets of the  $Z_4$. The most general Lagrangian, symmetric under $SU(3)\times SU(2)\times U(1)\times Z_4$,  contains the new terms 
\begin{align}\label{eq:L}
 \mathcal{L}&=\,\,\frac{1}{2}\mu_{S}^2S^2+\lambda_{S}S^4 +\frac{1}{2}\lambda_{S H}|H|^2S^2+M_{\psi}\overline{\psi}\psi+\frac{1}{2}\left[y_s\overline{\psi^c}\psi + y_{p}\overline{\psi^c}\gamma_5\psi + \rm{h.c.}\right]S, 
 \end{align}
where $H=[0, (h+v_H)/\sqrt{2}]^T$, with $h$ the SM Higgs boson. The mass of the real scalar singlet is then given by
\begin{align}
    M_{S}^2&=\mu_S^2+\frac{1}{2}\lambda_{SH}v_H^2.
\end{align}

From the Lagrangian one can see that $\psi$ is automatically stable whereas $S$ becomes stable for $M_S<2M_\psi$. In the following, this condition is assumed to hold so that both $S$ and $\psi$  contribute to the observed dark matter density. The model thus describes a two-component dark matter scenario. 

A couple of previous works have discussed similar scenarios in the past. Recently, a model without the $y_s$ term and with no $Z_4$ symmetry was  considered in Ref.~\cite{DiazSaez:2021pmg}. The structure of their fermion interaction term is, however, $\bar{\psi}\gamma^5\psi S$ rather than $\bar{\psi^c}\gamma^5\psi S$. Previously, in Ref.~\cite{Cai:2015zza}, a model based on the $Z_4$ symmetry and with the same particle content  was proposed, but the interaction term proportional to $y_p$ was left out and only few of its implications were studied.  A phenomenological analysis of the $Z_4$ model described above, including the impact of the most recent direct detection data and the characterization of its viable parameter space, is clearly due and is the goal of this work.

Even if it contains two species contributing to the dark matter, this $Z_4$ model is exceptionally minimal.  A single discrete symmetry stabilizes both dark matter particles, and five parameters ($M_S$, $M_\psi$, $\lambda_{SH}$, $y_s$, $y_p$) dictate the model phenomenology. It probably is the simplest model of two-component dark matter that can be envisioned, and it is  simpler that many of the standard (one-component) dark matter models that have been previously studied.  

Among the three new couplings, the Higgs-portal, $\lambda_{SH}$, plays a prominent role as it couples the dark matter sector with the SM particles. Notice that $\psi$ interacts directly only with $S$, which in turn couples to the Higgs and, through it, to the rest of the SM particles. Hence, $\lambda_{SH}$ must necessarily be different from zero, but either $y_s$ or $y_p$ can in principle vanish --not both though as $\psi$ would become a free particle. It will be convenient, in our analysis, to separately consider the cases $y_p=0$ and $y_s=0$, which we refer to as the scalar portal and the pseudoscalar portal respectively.  In this work, we will focus on the \emph{freeze-out} regime~\cite{Steigman:2012nb} of this model\footnote{Freeze-in production~\cite{Hall:2009bx} can also be realized.}, where the couplings are large enough for the dark matter particles to reach thermal equilibrium in the early Universe, and which typically leads to observable signals in dark matter experiments.   

This model can be seen as a merging of two (one-component) dark matter models that have been extensively studied in the literature: the singlet scalar~\cite{Silveira:1985rk,McDonald:1993ex,Burgess:2000yq} and the singlet fermion~\cite{Kim:2006af,Kim:2008pp, Lopez-Honorez:2012tov,Esch:2013rta}.  Both are highly constrained by current data stemming from the relic density and direct detection limits but, as we will show, these constraints can be greatly relaxed when we combine these two models into  the single two-component dark matter scenario described by equation (\ref{eq:L}). In fact, in our model there are novel dark matter processes that affect the relic density and open up new viable regions of the parameter space.

\begin{figure}[t]
\centering
\includegraphics[scale=0.9]{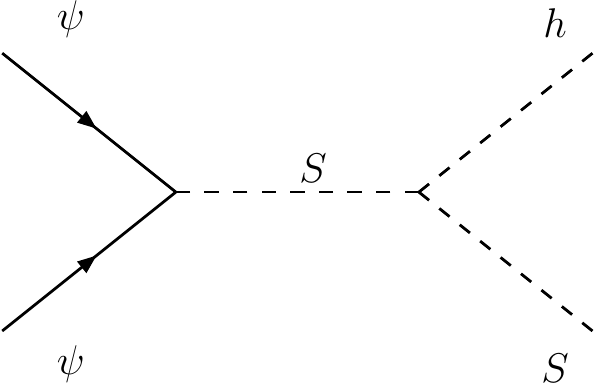}\hspace{0.4cm}
\includegraphics[scale=0.9]{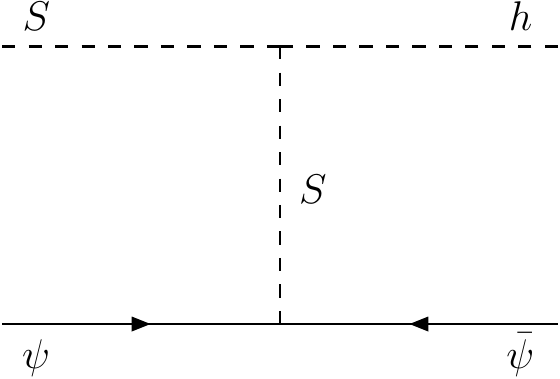}\hspace{0.4cm}\\
\vspace{0.4cm}
\includegraphics[scale=0.9]{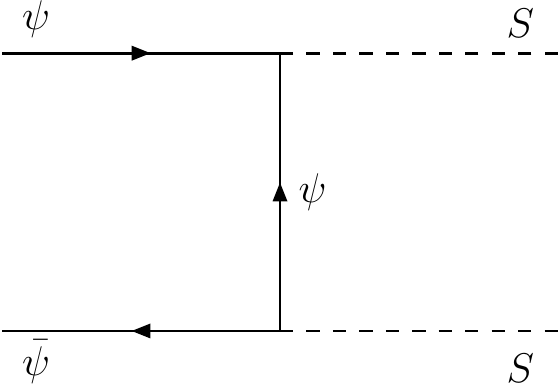}
\caption{Dark matter semiannihilation (top) and conversion (bottom) processes.}
\label{fig:semi-conv}
\end{figure}

\section{Dark Matter phenomenology}
\label{sec:DMpheno}

\subsection{Dark Matter processes}
The  terms in  $\mathcal{L}$ affect the dark matter phenomenology in different ways.
The interplay of the interactions controlled by $\lambda_{SH}$ and $y_{s,p}$  lead to $\psi\psi$ and $\psi S$ semiannihilations~\cite{Hambye:2008bq,DEramo:2010keq} (top panels in Figure \ref{fig:semi-conv}), while  the Yukawa interactions $y_s$ and $y_p$ lead to dark matter conversions  (bottom panel in Figure \ref{fig:semi-conv}). 
On the other hand, the Higgs portal interaction induces as usual scalar selfannihilations into a pair of fermions, weak gauge bosons and Higgses.    
At large $M_S$ the main annihilation channel is $SS\to hh$, with a cross section of the order of 
\begin{align}
    \sigma v(S S \to hh)\sim \frac{\lambda_{SH}^2}{16\pi M_S^2}.
\end{align}

\subsubsection{$\psi\psi$ annihilation}
The processes $\psi\psi\to Sh$ and $\bar\psi \bar\psi\to Sh$ generate a modification in the $\psi$ number density by two units (and in the $S$ number by one unit).  
The cross section for $\psi\psi\to Sh$ is given by
\begin{align}
    \sigma v(\psi\psi\to Sh) &= \, \frac{\lambda_{SH}^2v_H^2\beta(M_S,M_h)}{16\pi s^2(s-M_S^2)^2} \left[(s-4M_\psi^2)|y_s|^2 + s|y_p|^2 \right].
\end{align}
where
\begin{align}
    \beta(M_i,M_j)&=\left[s^2-2s(M_i^2+ M_j^2) + (M_i^2 - M_j^2)^2\right]^{1/2}.
\end{align}
Expanding it in terms of even powers of the relative velocity $v$ we obtain $\sigma v(\psi\psi\to Sh) = a_1+b_1v^2$ 
with 
\begin{align}
    a_1&=\frac{\sqrt{M_h^4+(M_S^2-4M_\psi^2)^2-2M_h^2(M_S^2+4M_\psi^2)}}{64\pi M_\psi^2(M_S^2-4M_\psi^2)^2}\lambda_{SH}^2 |y_p|^2,\\
    b_1& = \left(-C_p|y_p|^2+C_s|y_s|^2\right)\frac{\lambda_{SH}^2v_H^2}{\Delta}.
\end{align}
The expressions for $\Delta, C_p$ and $C_s$ are reported in the Appendix. 
This process becomes velocity suppressed for $y_p=0$ and the process is kinematically favourable as long as $2M_\psi> M_S+M_h$. 

Concerning the reverse process $S h \to \psi\psi$, the expression for $\sigma v(Sh\to \psi\psi)=\tilde{a}_1+\tilde{b}_1v^2$ at order $\mathcal{O}(v^0)$ is
\begin{align}
\tilde{a}_1&=  \frac{v_H^2 \lambda _{SH}^2 \sqrt{2 M_h M_S+M_h^2+M_S^2-4 M_{\psi }^2} }{32 \pi  M_h^3 M_S \left(M_h+M_S\right)
   \left(M_h+2 M_S\right){}^2}\left[((M_h + M_S)^2-4M_\psi^2)|y_s|^2 +(M_h + M_S)^2|y_p|^2\right].
\end{align}
Due to the relative minus sign present in the coefficient $((M_h + M_S)^2-4M_\psi^2)$ accompanying $|y_s|^2$, some interference effects are expected to occur in the resulting thermally averaged cross section, which can be enhanced when both portals are opened.   

\subsubsection{$\psi S$ semiannhilation }
The processes $\psi S\to \bar \psi h$ and $\bar\psi S\to \psi h$ generate a modification in the $S$ number density by one unit.  
The differential cross section can be cast as
\begin{align}
    \frac{d\sigma}{d\Omega} (\psi S \to \bar{\psi}h) &= \, \frac{\lambda_{SH}^2v_H^2\beta(M_\psi,M_h)}{32\pi^2 s \beta(M_\psi,M_S) (t-M_S^2)^2} \left[ (2M_\psi^2-t/2)|y_s|^2-t/2|y_p|^2 \right].
\end{align}
The corresponding cross section in terms of $v$ gives $\sigma v(\psi S\to \bar{\psi}h) = a_2+b_2v^2$ with
\begin{align}
    a_2 &= \kappa' \lambda_{SH}^2v^2_H \left[(M_S^2-M_h^2)|y_p|^2+((M_S+2M_\psi)^2-M_h^2)|y_s|^2\right],\\
    b_2 &= \frac{\lambda_{SH}^2v^2_H M_{\psi }^2 }{(M_S+M_\psi)^3\Delta'} \left[ \left(M_h^2-(M_S+2 M_{\psi })^2\right)C'_s|y_s|^2 + \left(M_h^2-M_S^2\right)  C'_p |y_p|^2\right],
\end{align}
and
\begin{align}
       \kappa'&=\frac{\sqrt{(M_S^2-M_h^2)[(M_S+2M_\psi)^2-M_h^2]}}{32\pi M_S(M_S+M_\psi)[M_S^3+M_\psi(2M_S^2-M_h^2)]^2}.
\end{align}
The expressions for $\Delta', C_p'$ and $C_s'$ are reported in the Appendix\footnote{We notice that these results are not in agreement with those reported in Ref.~\cite{Cai:2015zza}.}. 

\begin{figure}[t]
\centering
\includegraphics[scale=0.44]{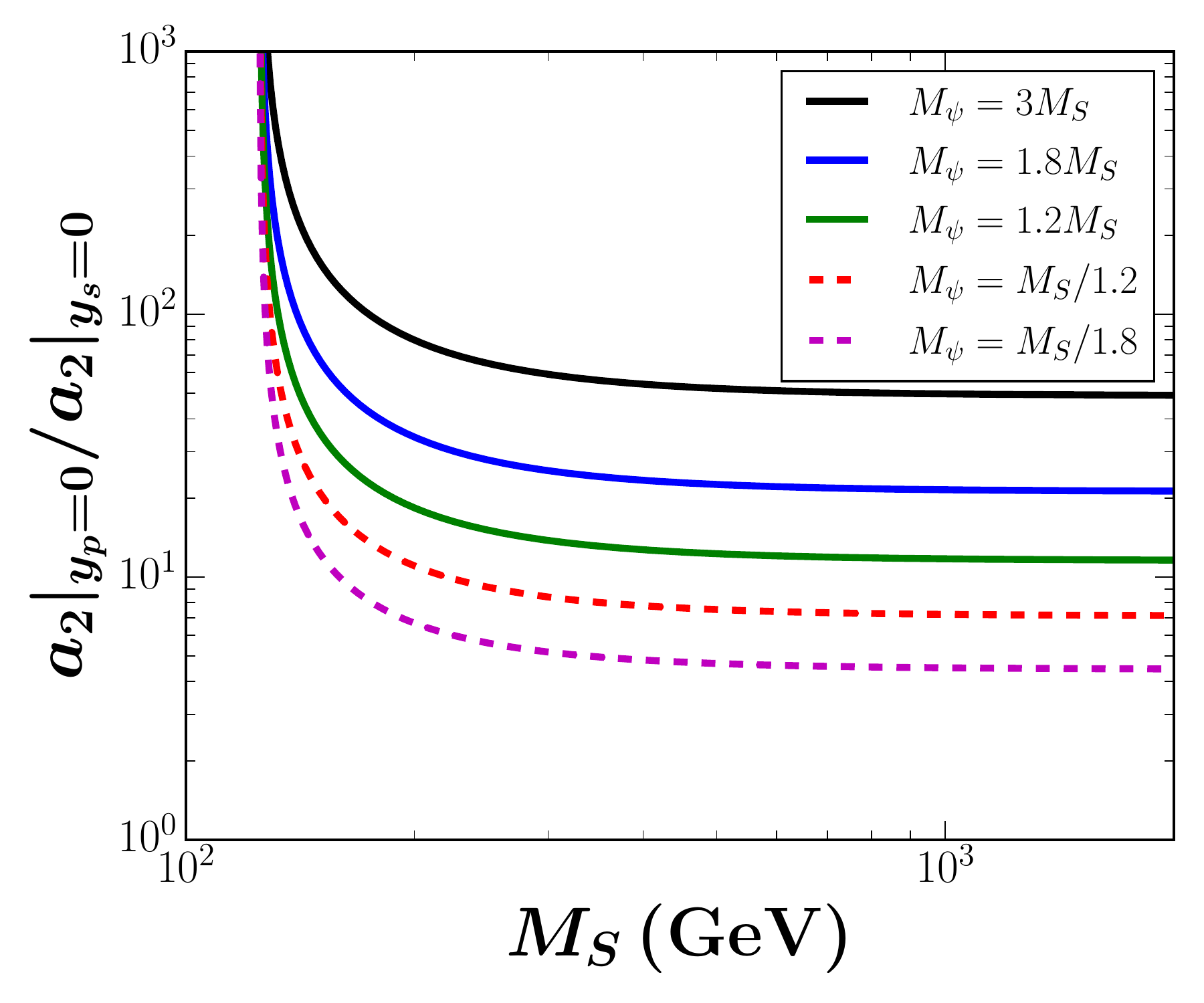}
\caption{$a_2|_{yp=0}/a_2|_{ys=0}$ as a function of $M_S$ for different mass ratios $M_\psi/M_S$. }  
\label{fig:a2ratio}
\end{figure}

This cross section does not suffer a velocity suppression in either case --$y_s=0$ or $y_p=0$. It receives, instead, an enhancement, in the case $y_p=0$, due to the dependence with $M_\psi$ in the velocity independent factor $a_2$, which strengthens the $S$ semiannihilation in comparison with the case $y_s=0$ (see figure \ref{fig:a2ratio}). For $M_S\gg M_h$ the ratio $a_2|_{yp=0}/a_2|_{ys=0}$ reaches the asymptotic value  $(1+2M_\psi/M_S)^2$.

Comparing the rates for the scalar selfannihilation and semiannihilation processes, the  former will dominate if
\begin{align}
   |y_s|&>\sqrt{\frac{2(1+M_\psi/M_S)}{(1+2M_\psi/M_S)}}\frac{M_S}{v_H},\\
   |y_p|&>\frac{2\sqrt{3}M_S}{v_H},
\end{align}
for the case of $y_p=0$ and $y_s=0$, respectively. Thus, the semiannihilation processes are typically efficient for not so large scalar masses, and in the $y_s=0$ case if $y_p\gtrsim1$ is also fulfilled. 

\subsubsection{$\bar \psi \psi \to SS$}
The differential cross section for $\bar \psi \psi \to SS$ is
\begin{align}
    \frac{d\sigma}{d\Omega} (\bar \psi \psi  \to SS) &= \,\frac{\beta(M_S,M_S)}{64\pi s \beta(M_\psi,M_\psi)}\left[\frac{\Sigma_t}{2(t-M_\psi^2)^2}+\frac{\Sigma_u}{2(u-M_\psi^2)^2}+\frac{\Sigma_{tu}}{(t-M_\psi^2)(u-M_\psi^2)}\right],
\end{align}
where the $\Sigma$ functions are reported in the Appendix. 
The corresponding cross section in terms of $v$ turns to be always velocity suppressed, in other words,  expressing $\sigma v(\bar \psi \psi \to SS) = a_3+b_3v^2$ implies that  
\begin{align}
    a_3 &= 0,\\
    b_3&=\frac{M_\psi\sqrt{M_{\psi }^2-M_S^2}}{24 \pi  \left(M_S^2-2 M_{\psi }^2\right)^4}\Delta_3,
\end{align}
with 
\begin{align}
    \Delta_3=&-2 M_S^2 M_{\psi }^2 \left(-y_p^2
   y_s^2+y_p^4+4 y_s^4\right)+M_S^4 \left(y_p^4+2 y_s^4\right)+M_{\psi }^4 \left(-2 y_p^2
   y_s^2+y_p^4+9 y_s^4\right).
\end{align}

\subsection{The Boltzmann equations}
\begin{table}[t]
        \begin{tabular}{c |c}
      $\psi$ Processes   & Type  \\
      \hline
        $\psi+\bar\psi \to S+S$  & $1122$\\
        $\psi+\psi\to S + h$ & $1120$\\
    \end{tabular}\hspace{2cm}
    \begin{tabular}{c |c}
      $S$ Processes   & Type  \\
      \hline 
        $S+S\to SM + SM$ & $2200$\\
        $S+S \to \psi+\bar\psi$  & $2211$\\
        $S+h \to \psi+\psi$  & $2011$\\
        $S+\psi \to \bar\psi+h$  & $2110$\\
    \end{tabular}
    \caption{The $2\to 2$ processes that are allowed (at tree-level) in the $Z_4$ model and that can modify the relic density of $\psi$ (left) and $S$ (right). $h$ denotes the SM Higgs boson. Conjugate and inverse processes are not shown.  
    }
    \label{tab:processesZ4}
\end{table}

The processes that may affect the $\psi$ and $S$ relic densities are summarized in Table \ref{tab:processesZ4}, and classified according to their type. For this classification, $\psi$ and $S$ are assumed to belong respectively  to sectors $1$ and $2$ while the SM particles belong to sector $0$. Notice, in particular, that processes of the type $1100$ are not allowed as $\psi$ cannot annihilate at tree-level into SM particles.  The Boltzmann equations for our model can then be written down as
\begin{align}
\label{boltzmann1}
\frac{dn_\psi}{dt}&= -\sigma_v^{1120}\left( n_\psi^2- n_S \frac{\bar{n}_\psi^2}{\bar{n}_S} \right)
- \sigma_v^{1122}\left( n_\psi^2- n_S^2 \frac{\bar{n}_\psi^2}{\bar{n}_S^2}
\right)  - 3H n_\psi, \\
\frac{dn_S}{dt}&=-\sigma_v^{2200}  \left(n_S^2-\bar{n}_S^2 \right) 
- \sigma_v^{2211}\left( n_S^2- n_\psi^2 \frac{\bar{n}_S^2}{\bar{n}_\psi^2}
\right)-\frac{1}{2}\sigma_v^{1210}\left( n_\psi n_S- n_\psi \bar{n}_S \right)\nonumber\\
&~~+\frac{1}{2}\sigma_v^{1120}(n_\psi^2-n_S\frac{\bar{n}_\psi^2}{\bar{n}_S})   - 3H n_S.      
\label{boltzmann2}
\end{align}
Here $\sigma_v^{abcd}$  stands for  the thermally averaged cross section, which satisfies 
\begin{equation}
    \bar{n}_a\bar{n}_b\sigma_v^{abcd}=\bar{n}_c\bar{n}_d\sigma_v^{cdab},
\end{equation}
whereas $n_{\psi,S}$  denote the number densities of $\psi,S$, and $\bar{n}_{\psi,S}$  their respective equilibrium values.  To numerically solve these equations and obtain the relic densities, we rely on {\tt micrOMEGAs}~\cite{Belanger:2014vza} throughout this paper. Since
 its version 4.1, {\tt micrOMEGAs} incorporated two-component dark matter scenarios, automatically taking into account all the relevant processes in a given model.
 
 \begin{figure}[t]
\centering
\includegraphics[scale=0.44]{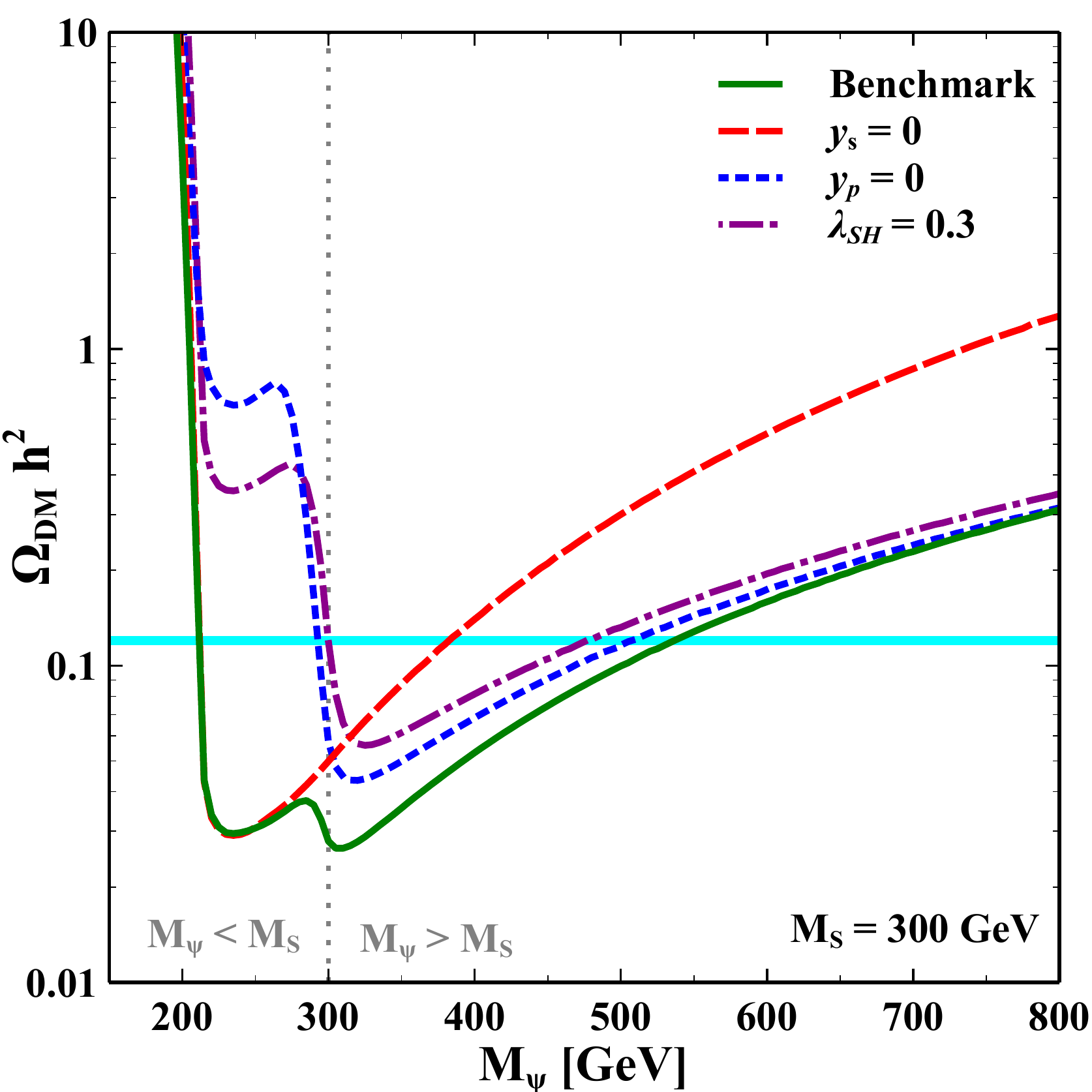}
\includegraphics[scale=0.44]{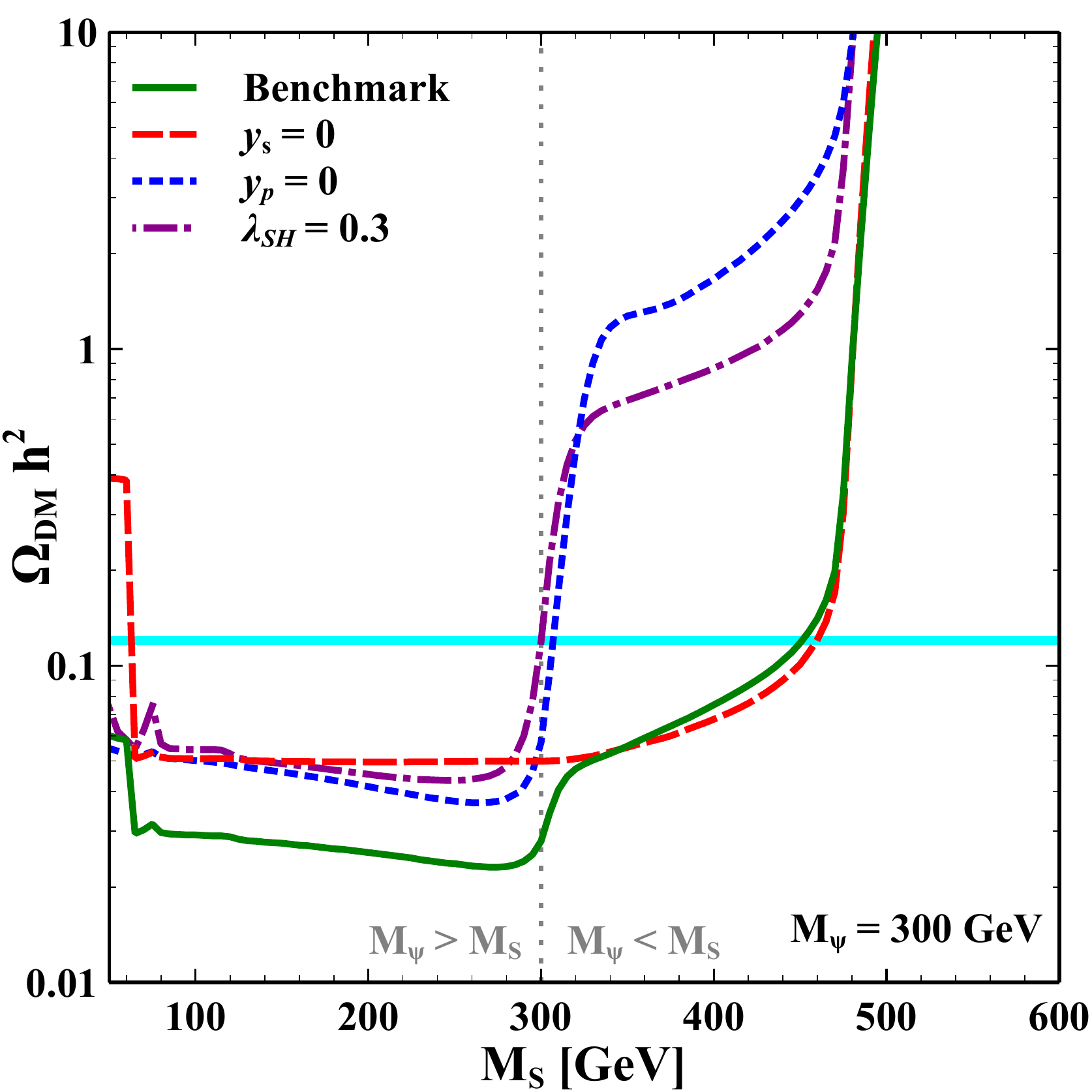}\\
\caption{The total relic density as a function of $M_\psi$ (left) or $M_S$ (right) for different set of parameters. In the left (right) panel $M_S$ ($M_\psi$) is fixed to $300~$ GeV. The benchmark model, in solid green,  features $y_s=y_p=\lambda_{SH}=1$. The other lines differ from the benchmark only on the value of the coupling shown in the key.}  
\label{fig:relics}
\end{figure}

\begin{figure}[t]
\centering
\includegraphics[scale=0.44]{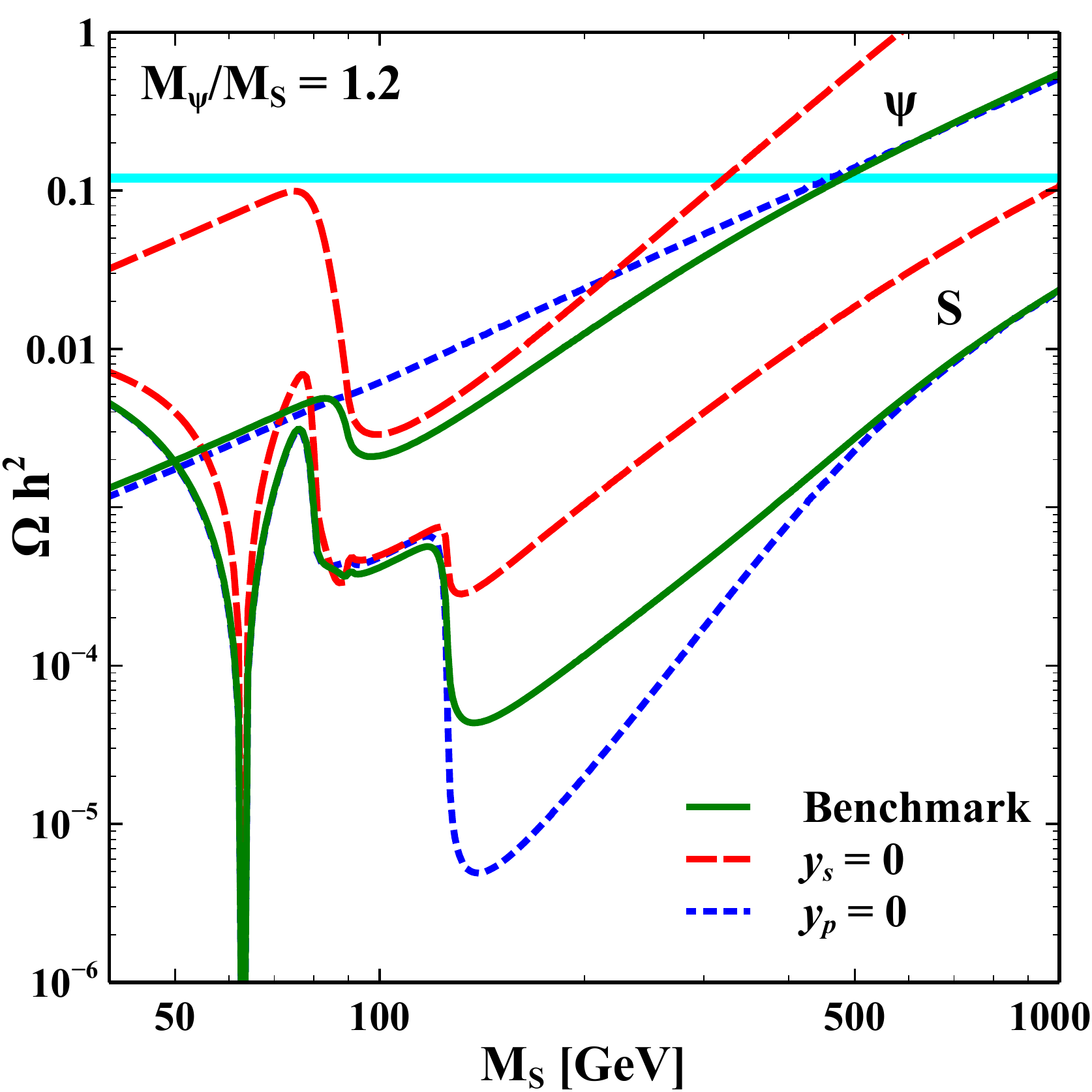}
\includegraphics[scale=0.44]{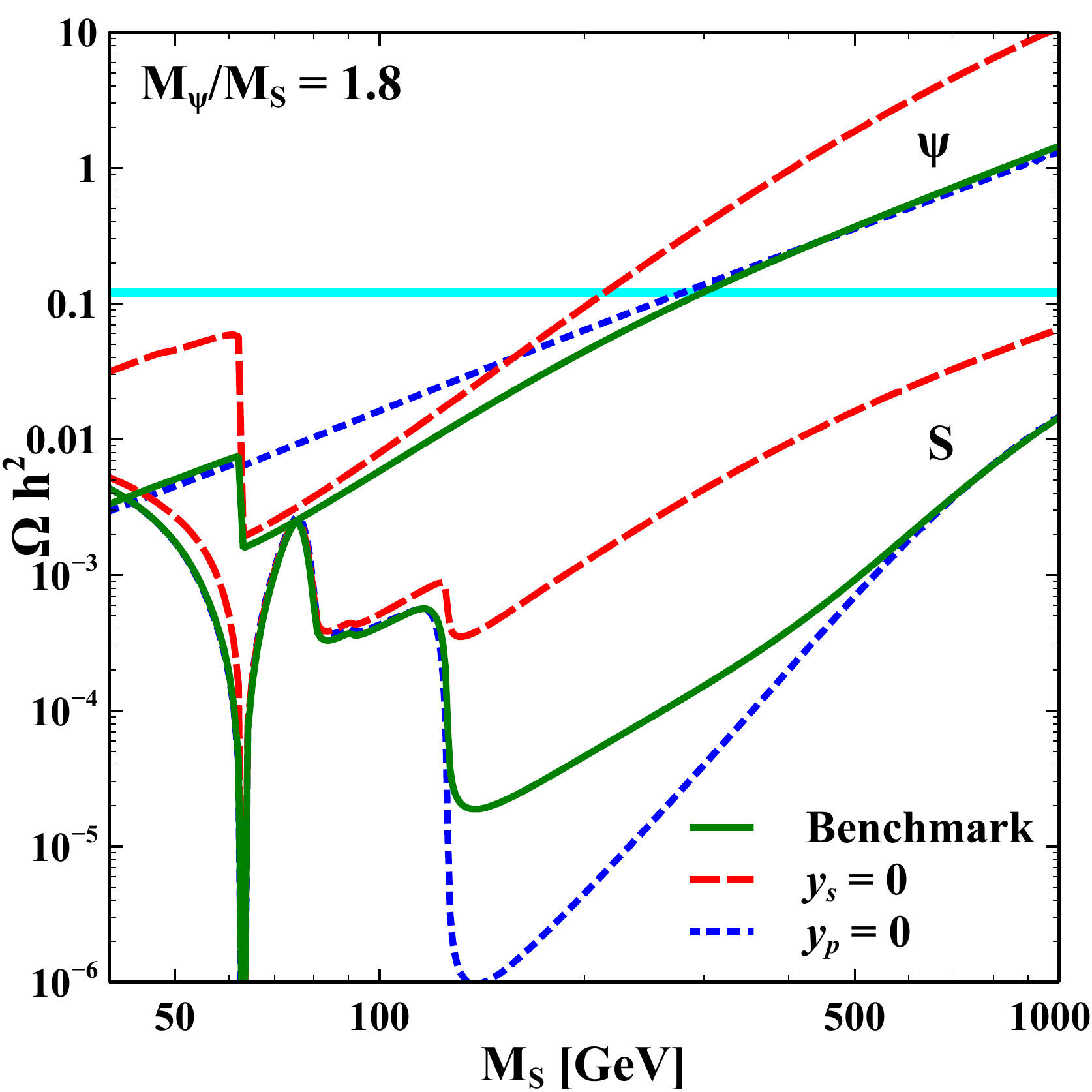}\\
\caption{The $\psi$ and $S$ relic densities as  functions of  $M_S$ for $M_\psi/M_S=1.2$ (left) and $M_\psi/M_S=1.8$ (right).  The benchmark model, in solid green,  features $y_s=y_p=\lambda_{SH}=1$. The other lines differ from the benchmark only on the value of the coupling shown in the key.}  
\label{fig:ratios12}
\end{figure}

\begin{figure}[t]
\centering
\includegraphics[scale=0.44]{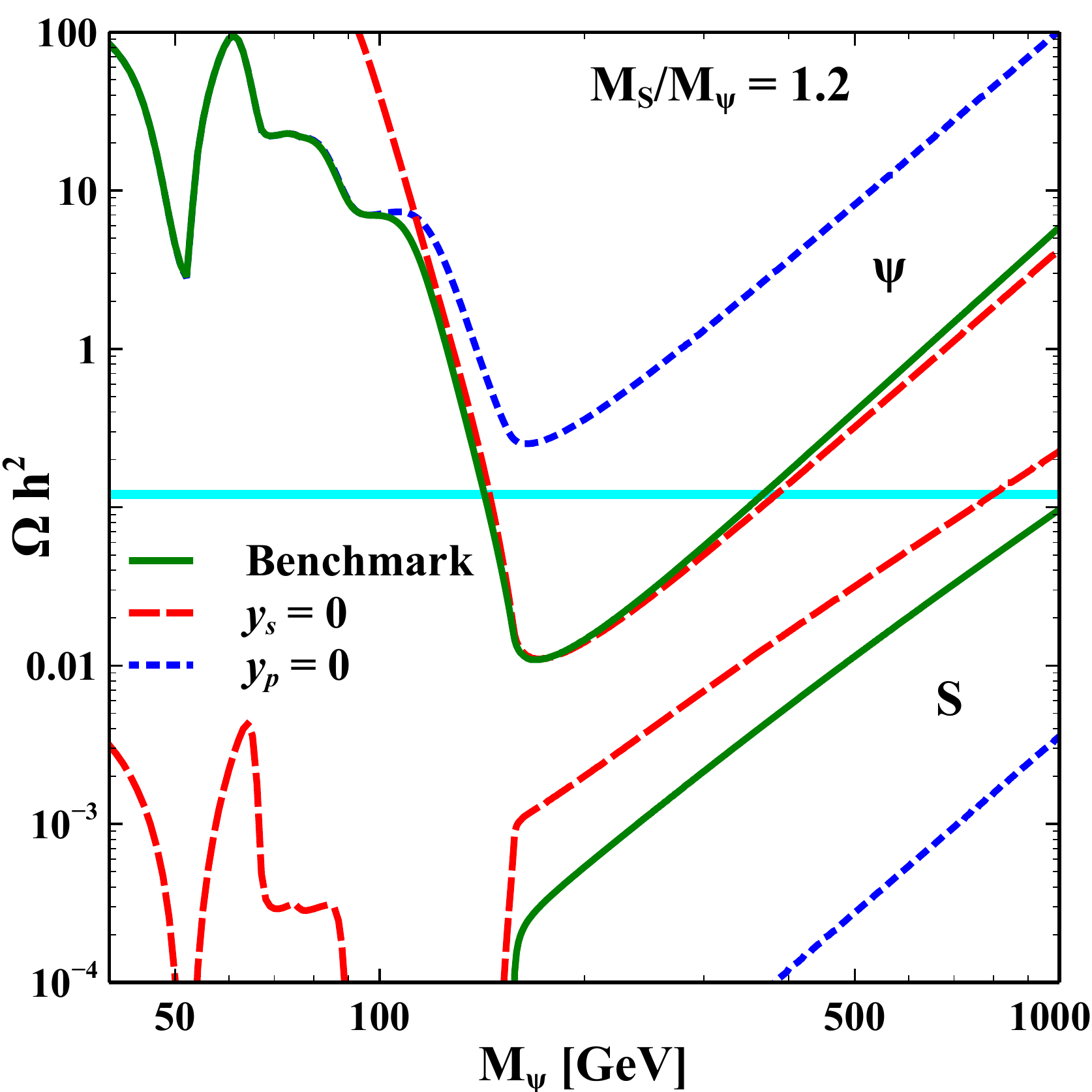}
\includegraphics[scale=0.44]{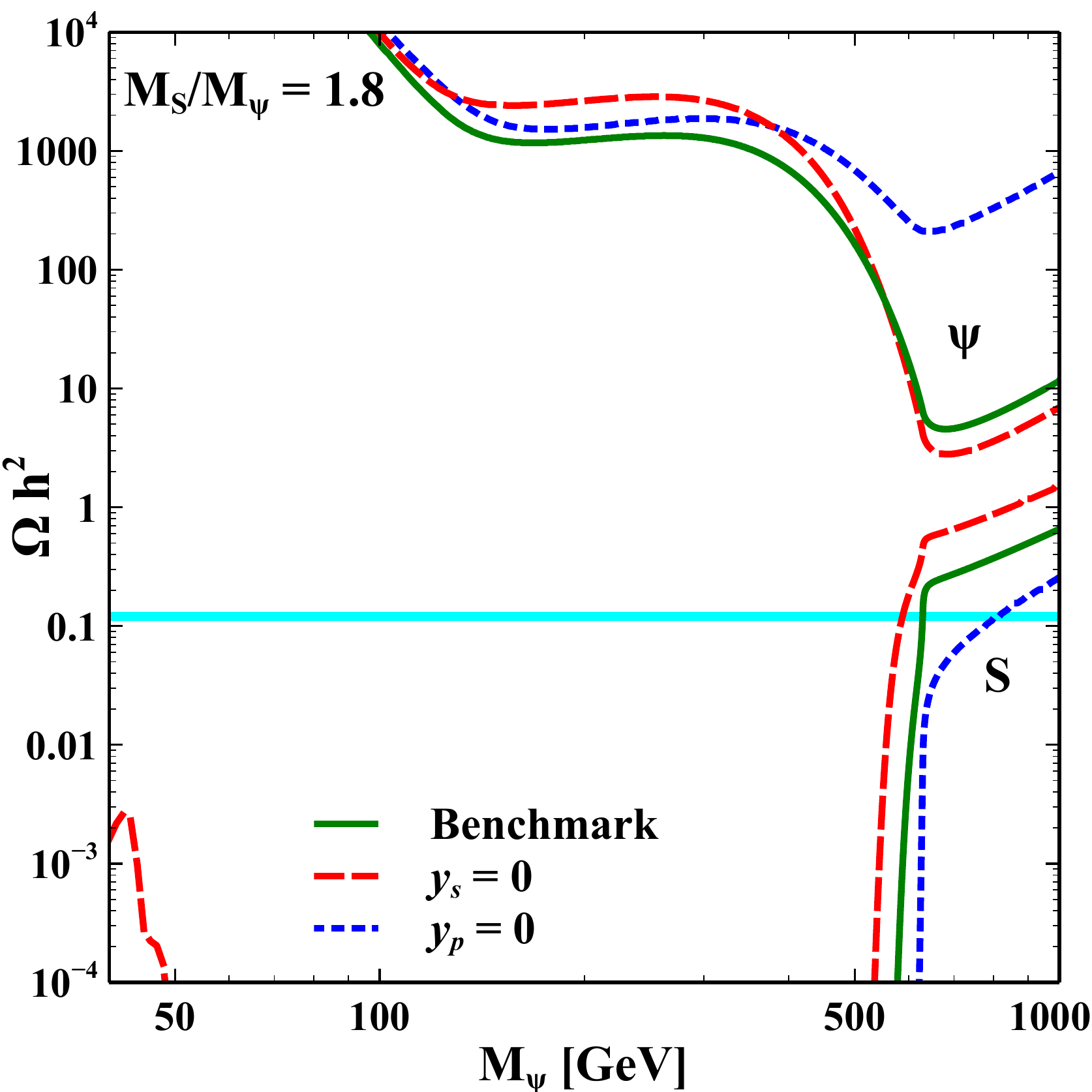}\\
\caption{The $\psi$ and $S$ relic densities as  functions of  $M_\psi$ for $M_S/M_\psi=1.2$ (left) and $M_S/M_\psi=1.8$ (right).  The benchmark model, in solid green,  features $y_s=y_p=\lambda_{SH}=1$. The other lines differ from the benchmark only on the value of the coupling shown in the key.}  
\label{fig:ratios34}
\end{figure}

 To illustrate the solutions to the Boltzmann equations in our model, figure \ref{fig:relics} shows the total relic density, $\Omega_\psi+\Omega_S$, for four diverse sets of couplings. In the benchmark model (green solid line), all three couplings are equal to one: $\lambda_{SH}=y_s=y_p=1$; the other lines differ from the benchmark only on the value of \emph{one} coupling, which is specified in the key. Thus, the dashed blue line, for instance,  corresponds to $\lambda_{SH}=y_s=1$ and $y_p=0$. In the left panel, we set $M_S=300$ GeV and vary $M_\psi$, whereas in the right panel the roles of $M_S$ and $M_\psi$ are exchanged. The vertical (gray) dotted line separates the two possible mass regimes in this model: $M_S<M_\psi$ and $M_S>M_\psi$. Since $M_S<2M_\psi$ (to ensure a two-component dark matter scenario),  in the left panel the minimum allowed value of $M_\psi$ is $150$ GeV, whereas in the right panel the maximum possible value of $M_S$ is $600$ GeV. The horizontal (cyan) band represents the observed valued of the dark matter density. From this figure, we can already see that it is possible to satisfy the dark matter constraint in both mass regimes and for different values of the couplings. To better understand the behavior observed in this figure, it is necessary to look separately at the relic densities of $\psi$ and $S$, as done in figures  \ref{fig:ratios12} and \ref{fig:ratios34}.

 Figure \ref{fig:ratios12} displays the relic densities of $\psi$ (upper lines) and $S$ (lower lines) as a function of $M_S$ for three sets of couplings. The difference between the two panels is the value of $M_\psi/M_S$: $1.2$ (left) and $1.8$ (right) --both corresponding to the regime $M_S<M_\psi$. The $S$ relic density has the well-known shape of the singlet scalar model (the Higgs resonance is clearly visible) up to $M_S\sim M_h$, where the semiannihilation process $S+\psi\to \bar\psi+h$ becomes kinematically allowed. The semiannihilations are more efficient in decreasing $\Omega_S$ for $y_p=0$ than for $y_s=0$ as expected (see figure~\ref{fig:a2ratio}). 
 The $\psi$ relic density instead drops, for the benchmark and for $y_s=0$, around $M_S=90$ GeV, where the process $\psi+\psi\to S+h$ starts contributing to the annihilation rate. For $y_p=0$ (dotted blue line) this  process is velocity suppressed and its effect on the relic density  becomes negligible, being driven by the dark matter conversion processes. Notice that  the relic densities for the benchmark and the $y_p=0$ tend to converge at high masses (where the annihilations via the Higgs portal are the dominant ones) while differing from the $y_s=0$ case. 
For the higher value of $M_\psi/M_S$ the behavior of the relic densities is qualitatively similar.  In particular,  the fermion relic densities are always larger than the scalar ones.

Figure \ref{fig:ratios34} is analogous to \ref{fig:ratios12} but for the other mass regime, $M_\psi<M_S$. Two important differences appear in this case. For the fermion, the dark matter conversion process, $\psi+\bar\psi\to S+S$, is now kinematically suppressed (more so in the right panel) so that the only efficient way to reduce the $\psi$ density is via the semiannihilation process, $\psi+\psi\to S+h$. This process is allowed for $M_\psi\gtrsim 156$ GeV (left panel) and for $M_\psi\gtrsim 625$ GeV (right panel), explaining the change of behavior observed in the figures. For the scalar, there can be an exponential suppression of the relic density induced by the process $S+h\to \psi\psi$. This exponential behavior is rather common is multi-component dark matter scenarios and had already been observed in other models~\cite{Belanger:2020hyh}.  From the figure it is seen to be particularly relevant for high values of $M_S/M_\psi$ (right panel). 
By comparing the two panels, it is seen that the relic densities are higher the larger $M_S/M_\psi$ is. In the right panel, in fact,  the $\psi$ relic density lies well above the observed value over the entire range of $M_\psi$ and for all three sets of couplings, suggesting that  the dark matter constraint is more easily satisfied for small values of $M_S/M_\psi$.  Note also that, as before, the fermion relic density tends to be larger than the scalar one --a result that will be confirmed by our numerical analysis.
 
Besides the relic density, the parameter space of this model is  significantly restricted by direct detection limits, to which we now turn.

\subsection{Direct detection}

\begin{figure}[t]
\centering
\includegraphics[scale=1]{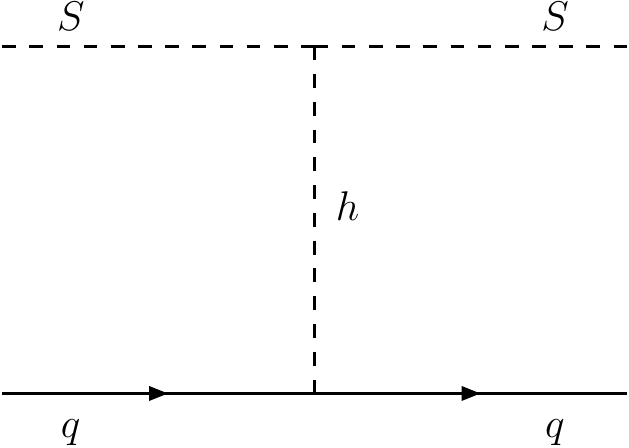}\hspace{1cm}
\includegraphics[scale=1]{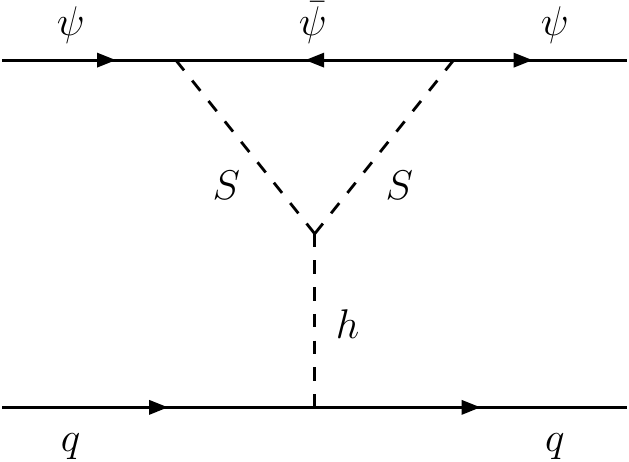}
\caption{Diagrams leading to the elastic scattering of dark matter particles off nuclei at one-loop level for the fermion (left panel) and at tree-level for the scalar (right panel).}  
\label{fig:DD-diagrams}
\end{figure}

As is common in dark matter models with scalar singlets, the elastic scattering of the dark matter particles off nuclei are possible thanks to the Higgs portal interaction $\lambda_{SH}$ (right panel of figure \ref{fig:DD-diagrams}). 
The expression for the spin-independent (SI) cross-section reads
 \begin{align}\label{eq:SIcxS}
     \sigma_{S}^{{\rm SI}}&=\frac{\lambda_{SH}^2}{4\pi}\frac{\mu_R^2 m_p^2 f_p^2}{m_h^4 M_{S}^2},
 \end{align}
 where $\mu_R$ is the reduced mass, $m_p$ the proton mass and $f_p\approx 0.3$ is the quark content of the proton. Because we are dealing with a two-component dark matter model, the relevant quantity to be compared against the experimental limits is, however, not $\sigma_S^{SI}$ itself but rather $\frac{\Omega_{S}}{\Omega_{DM}}\sigma_S^{SI}$, which takes into account the fact that $S$ contributes only a fraction of the observed dark matter density --the rest being due to $\psi$.

At tree level, $\psi$ cannot scatter elastically off nuclei, but it will do so at higher orders.  The one-loop diagram, which is expected to be the dominant contribution, is shown in the right panel of figure \ref{fig:DD-diagrams}. Even if loop-suppressed, this process will turn out to be within the sensitivity of current and future direct detection experiments, due to the significant values for $y_s$, $y_p$ and $\lambda_{SH}$ that are required to  annihilate $\psi$. The corresponding SI cross section  is given by  
 \begin{align}\label{eq:SIcxPsi}
     \sigma_{\psi}^{{\rm SI}}&=\frac{1}{\pi}\frac{\mu_{\psi p}^2 m_p^2 f_p^2}{m_h^4}\left[\lambda_{SH}\frac{|y_s|^2f(r_{S\psi})+|y_p|^2g(r_{S\psi})}{16\pi^2M_\psi}\right]^2,
 \end{align}
where $r_{S\psi}=M_S^2/M_\psi^2$ and 
\begin{align}
    f(r)&=\frac{r^2- 5 r+4}{\sqrt{(4-r) r}}\arctan\left(\frac{\sqrt{4-r}}{\sqrt{r}}\right)+ \frac{1}{2}[2-(r-3) \log (r)],\\
    g(r)&=\frac{(r-3) \sqrt{r}}{\sqrt{4-r}}\arctan \left(\frac{\sqrt{4-r}}{\sqrt{r}}\right)+\frac{1}{2}[2-(r-1) \log(r)].   
\end{align}
It is worth mentioning that the pseudo-scalar portal $y_p$ lead to a non velocity suppressed SI cross section. In contrast, the contribution  proportional to the product $y_sy_p$ has been neglected since it is suppressed by the square of the dark matter velocity (the corresponding direct detection bounds become weaker). Notice that in the limit $y_p=0$ the expression for $\sigma_{\psi}^{{\rm SI}}$ differs from  that reported in Ref.~\cite{Cai:2015zza}.

 We expect important restrictions on the viable parameter space of this model arising from direct detection limits, which should be imposed on those points satisfying the relic density constraint.  In the next two sections, we will randomly sample the five-dimensional parameter space of this model so as to obtain a large set of models compatible with all current data, including direct detection bounds. To facilitate the analysis, we will first study the regime $M_\psi<M_S$ and then switch to $M_S<M_\psi$. 
 
\section{The $M_\psi<M_S$ regime}\label{sec:pheno}
In this and the next sections,  we will obtain and analyze  viable regions for our two-component dark matter model.  To that end, the parameter space will first be randomly scanned, and the points compatible with all current bounds will be selected.  Our selection criteria include the constraints obtained from the invisible decays of the Higgs boson, the dark matter density~\cite{Aghanim:2018eyx} and direct dark matter searches~\cite{Aprile:2018dbl} --indirect dark matter searches do not significantly restrict the parameter space, as will be shown.  The resulting  sample of viable points will then be characterized, paying special attention to the appearance of new viable regions and  to the prospects for dark matter detection.  Let us emphasize that this random sampling of the parameter space does not warrant a statistical interpretation of the distribution of viable points (it cannot be used to find the most favored regions or the best fit points), but it will  help us to pinpoint the most relevant parameters and to identify the mechanisms that allow to satisfy the current bounds, which are our main goals. 
 
If $S$ is lighter than half the Higgs mass, the decay $h\to SS$ would be allowed,  contributing to the invisible branching ratio of the Higgs boson ($\mathcal{B}_{inv}$). The decay width associated with $h\to SS$ is
\begin{align}
    \Gamma(h\to SS)&=\frac{\lambda^2_{SH}v_H^2}{32\pi M_h}\left[1-\frac{4M^2_{S}}{M_h^2}\right]^{1/2}.
\end{align}
To be consistent with current data, we require that $\mathcal{B}_{inv}\leq0.13$) \cite{Sirunyan:2018owy,ATLAS:2020cjb}.

The relic density constraint reads 
\begin{equation}
    \Omega_{\psi}+\Omega_{S}=\Omega_{\text{DM}},
\end{equation}
where $\Omega_{\text{DM}}$ is the dark matter abundance as reported by PLANCK~\cite{Aghanim:2018eyx}, 
\begin{align}
    \Omega_{\text{DM}}h^2=0.1198\pm 0.0012. 
\end{align}
We consider a model to be compatible with this measurement if its relic density, as computed by micrOMEGAs, lies between $0.11$ and $0.13$, which takes into account an estimated theoretical uncertainty of order $10\%$. Since we have two dark matter particles, an important  quantity  in our analysis is the fractional contribution of each to the total dark matter density, $\xi_{\psi,S}\equiv \Omega_{\psi,S}/\Omega_{\text{DM}}$, with $\xi_{\psi} +\xi_{S}=1$.

Regarding direct detection, we require the spin-independent cross section, computed from equations (\ref{eq:SIcxS}) and (\ref{eq:SIcxPsi}), to be below the direct detection limit set by the XENON1T collaboration \cite{Aprile:2018dbl}. Such direct detection limit usually provides very strong constraints on Higgs-portal scenarios like the model we are discussing.  In particular, for the singlet real scalar  model~\cite{Silveira:1985rk,McDonald:1993ex,Burgess:2000yq} the minimum dark matter mass compatible with upper limit set by the XENON1T collaboration is $\sim950$ GeV (for the complex case turns to be $\sim2$ TeV). As we will show, however, the new interactions present in our two-component dark matter model permit to simultaneously satisfy the relic density constraint and direct detection limits for lower dark matter masses.

We will also study the testability of the viable  models at future direct detection experiments including  LZ~\cite{Akerib:2018lyp} and DARWIN \cite{Aalbers:2016jon}, as well as the possible constraints and the expected prospects from indirect detection searches. For these searches, the relevant particle physics quantity is no longer $\langle\sigma v\rangle$ but $\xi_i\xi_j\langle\sigma v\rangle_{ij}$, where  $\langle\sigma v\rangle_{ij}$ is the  cross section times velocity for the annihilation process of dark matter particles $i$ and $j$ into a certain final state. We will rely, on the theoretical side, on the computation of the different annihilation rates provided by   micrOMEGAS and, on the experimental side, on the limits and the projected sensitivities reported by the Fermi collaboration from  $\gamma$-ray observations of dShps  \cite{Ackermann:2015zua,Charles:2016pgz}.

In our scans the parameters are randomly chosen (using a logarithmically-uniform distribution) according to
\begin{align}
    &50\,{\rm GeV}\leq M_{\psi}\leq 2\,{\rm TeV},\,\,\,M_S<2M_\psi, \\
    &10^{-3}\leq |\lambda_{SH}|\leq 3,\\
    &10^{-2}\leq |y_s|,|y_p|\leq 3.
\end{align}

To better understand the role of the different parameters, the analysis will be divided into three cases: the scalar portal ($y_p=0$), the pseudoscalar portal ($y_s=0$), and the general case ($y_p, y_s\neq 0$). 

\subsection{Scalar portal}
\begin{figure}[t!]
\centering
\includegraphics[scale=0.4]{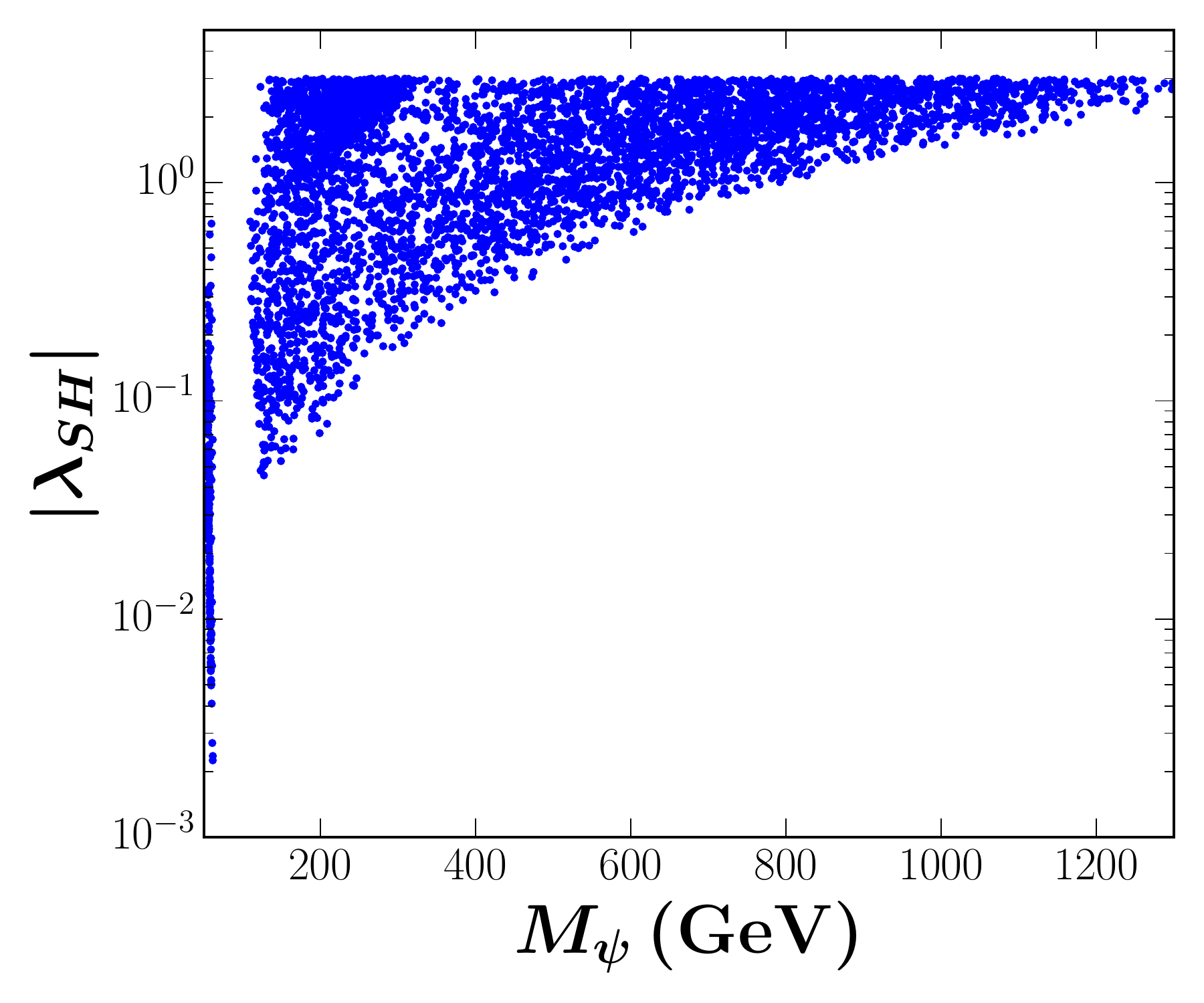}
\includegraphics[scale=0.4]{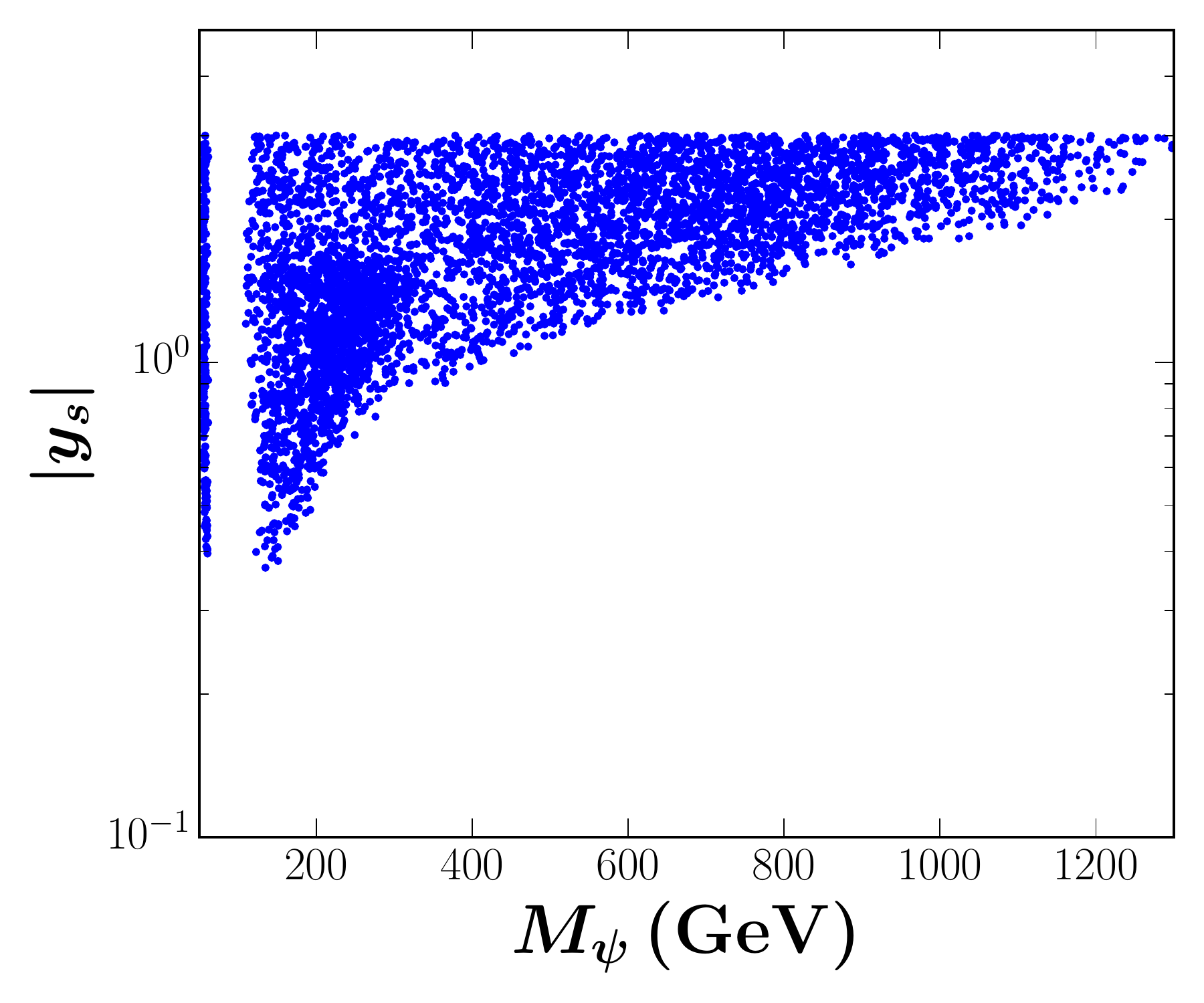}\\
\includegraphics[scale=0.4]{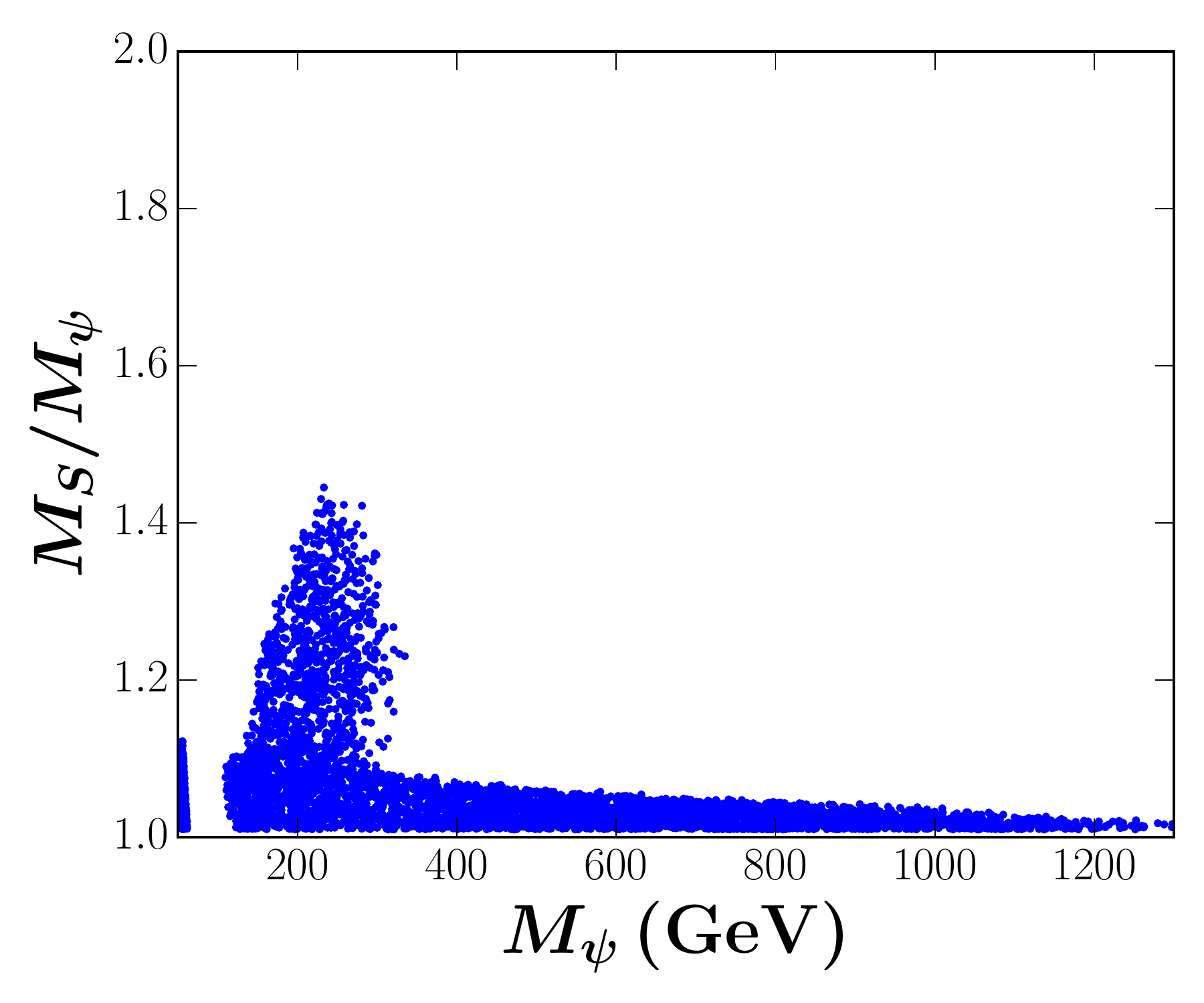}\\
\includegraphics[scale=0.4]{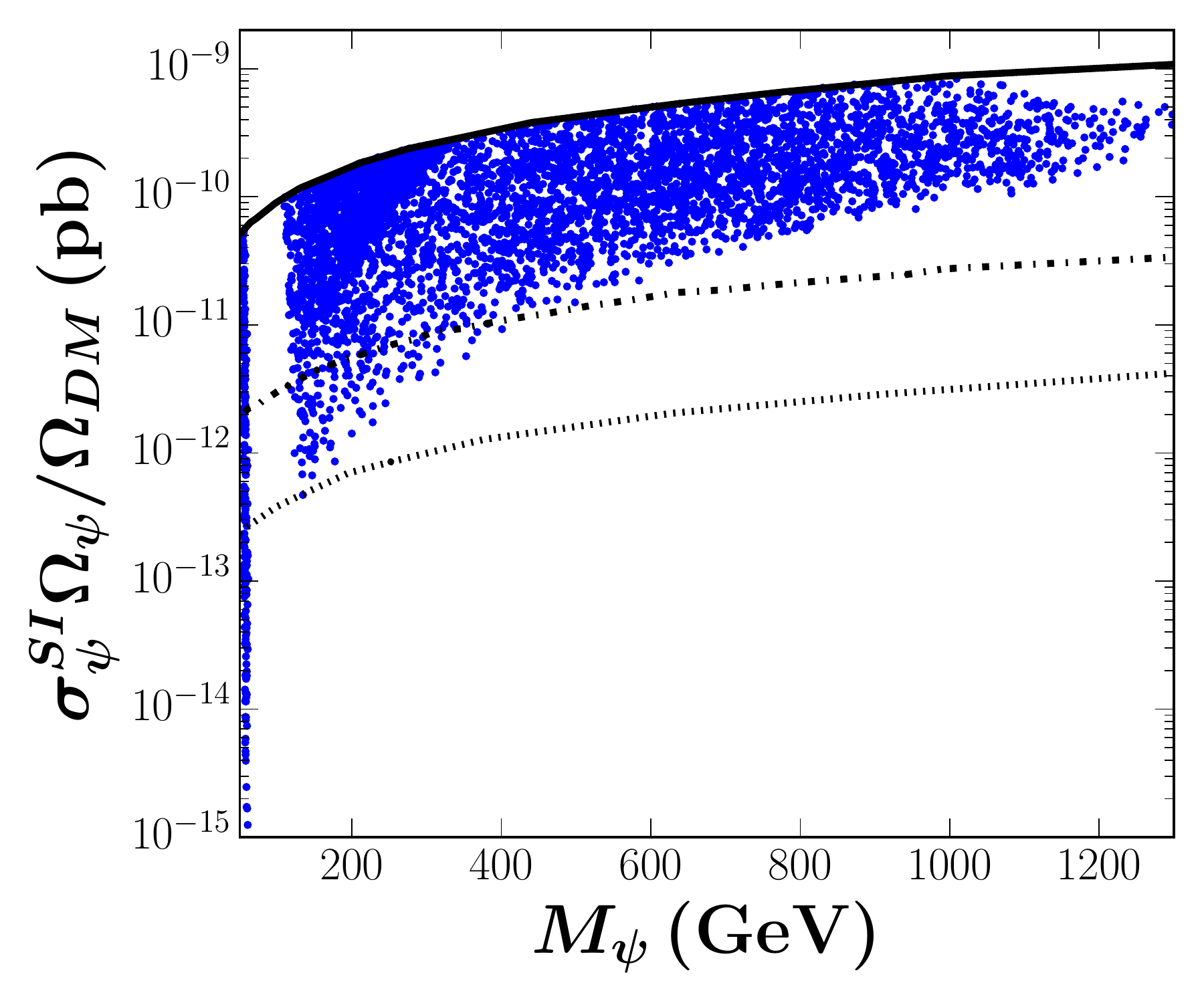}
\includegraphics[scale=0.4]{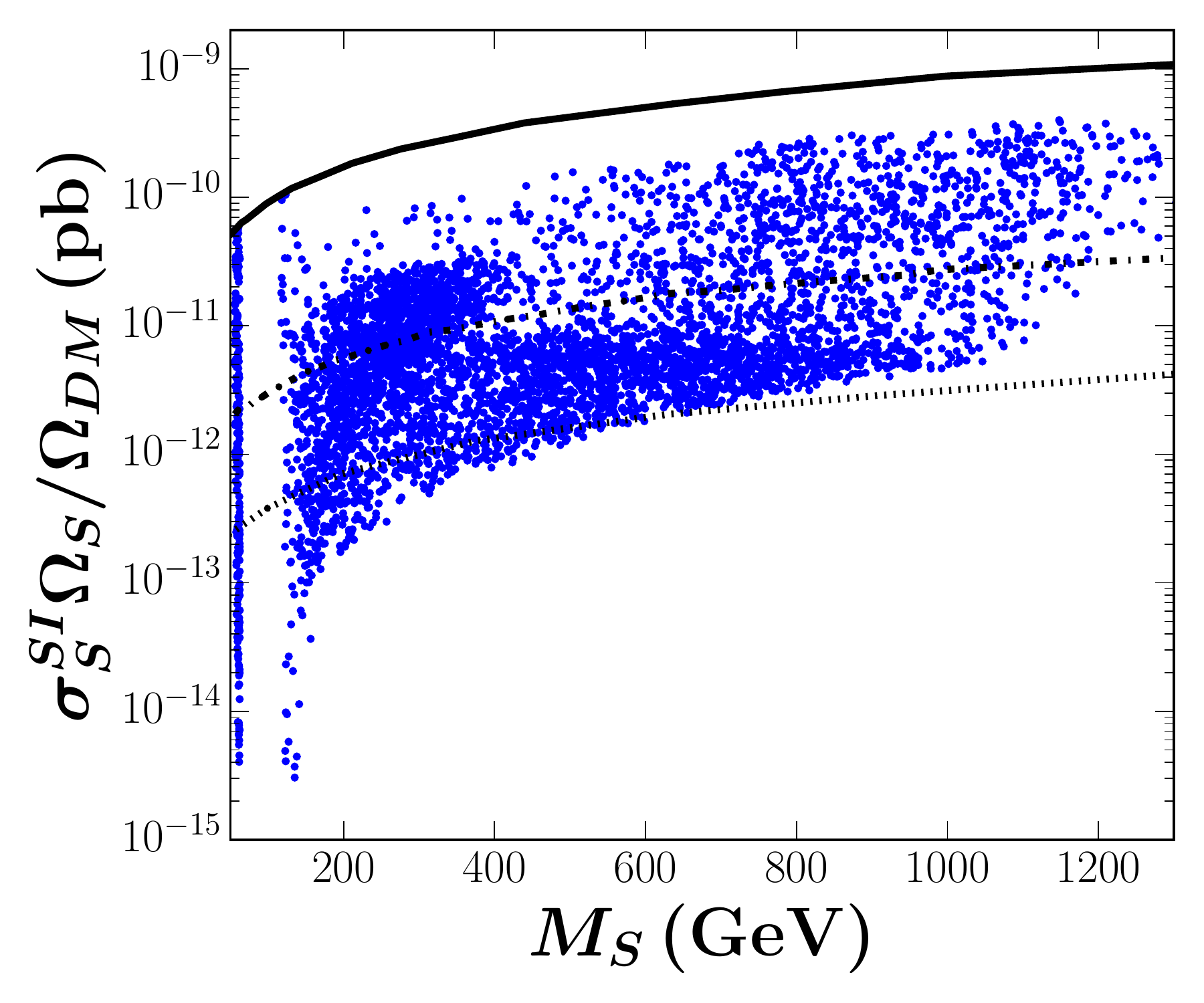}
\caption{A sample of viable models for $M_\psi<M_S$ and $y_p=0$ (scalar portal) projected onto different planes. The different panels show the couplings $\lambda_{SH}$ (top-left) and $y_s$ (top-right), the ratio of dark matter masses (center), and the direct detection prospects of $\psi$ (bottom-left) and $S$ (bottom-right). In the bottom panels, the lines correspond, from top to bottom, to the current limit from XENON1T, and the expected sensitivities of LZ and DARWIN.}  
\label{fig:ypnull}
\end{figure}

Here we have $y_p=0$ so that $\sigma v(\psi\psi\to Sh)$  and $\sigma v(\bar \psi \psi \to SS)$ are velocity suppressed. Figure \ref{fig:ypnull} displays a sample of viable models projected onto different planes. First of all, notice from the different panels that the viable models cover the entire range of dark matter masses below 1.3 TeV or so --an important result that demonstrate the existence of new viable regions not present in the singlet scalar or the singlet fermion models. From the top panels we see that, as expected,  $|\lambda_{SH}|$ and $|y_s|$ tend to  increase with $M_\psi$, reaching their maximum allowed value for $M_\psi\sim 1.3$ TeV. Higher dark matter masses would require couplings larger than  allowed in our scans. In the center panel,  two regions can be distinguished: $M_\psi\lesssim 400$ GeV, where the ratio $M_S/M_\psi$ can vary up to $1.5$ (the maximum is $2$) and semiannihilations play a crucial role, as shown later; and $M_\psi\gtrsim 400$ GeV, where $M_S/M_\psi$ is at most $1.1$ and the two dark mater particles become more degenerate with increasing mass.  In this region,  $\psi+\psi\to S+S$ is the key process that reduces the $\psi$ relic density, explaining why a mass degeneracy is required ($M_\psi>M_S$). The bottom panels compare the predicted elastic scattering rate off nuclei against the current limit (solid) and the expected sensitivities of planned experiments for the fermion (left) and the scalar (right). Notice that, for both dark matter particles, most of the viable points in our sample lie within the expected sensitivity of DARWIN. And for the fermion, most of them lie within the reach of LZ\footnote{In \cite{Cai:2015zza} it was instead found that the fermion contribution was always negligible.}. Direct detection experiments, therefore, constitute a very promising way to probe this scenario.      

\subsection{Pseudoscalar portal}
\begin{figure}[t!]
\centering
\includegraphics[scale=0.4]{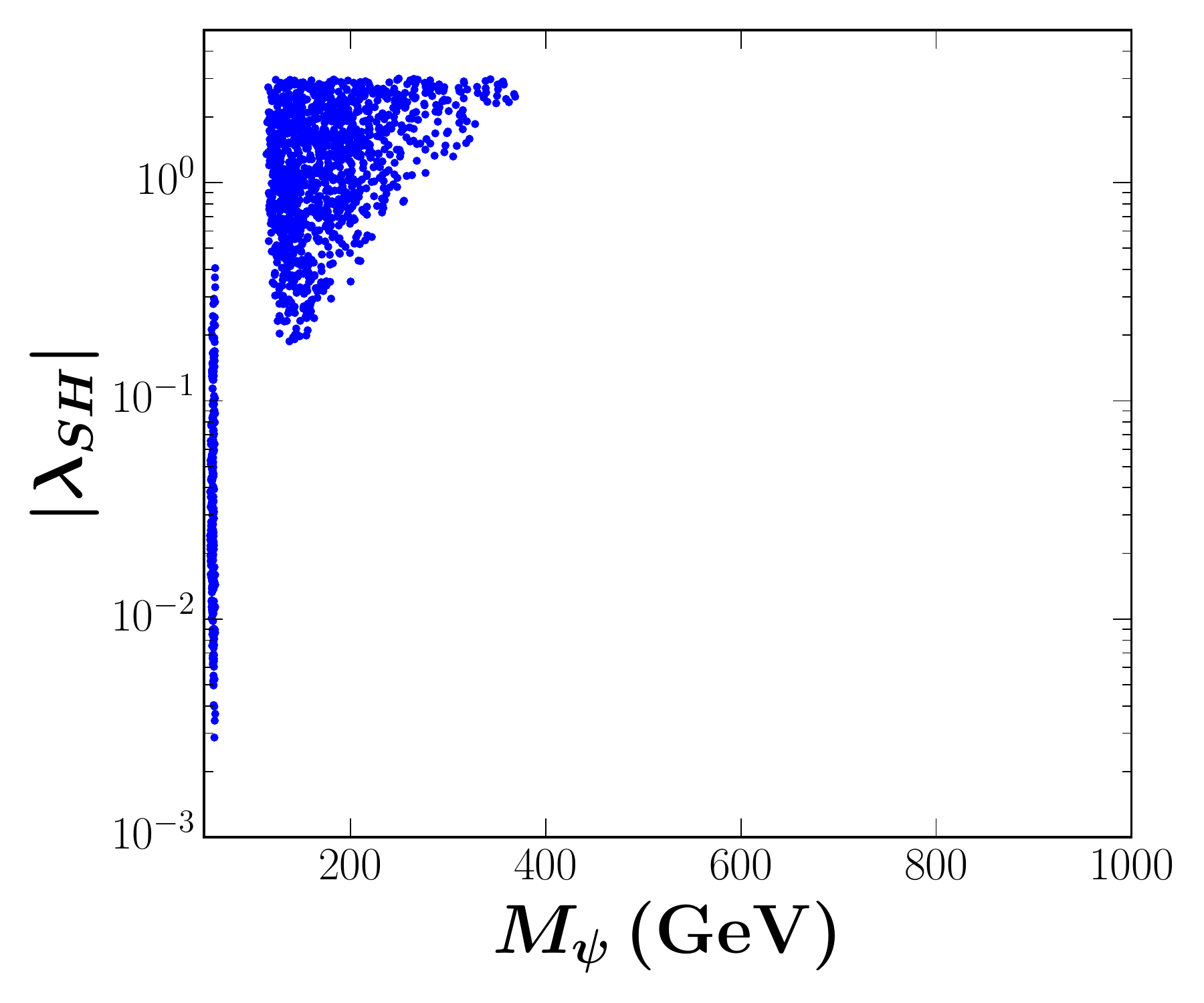}
\includegraphics[scale=0.4]{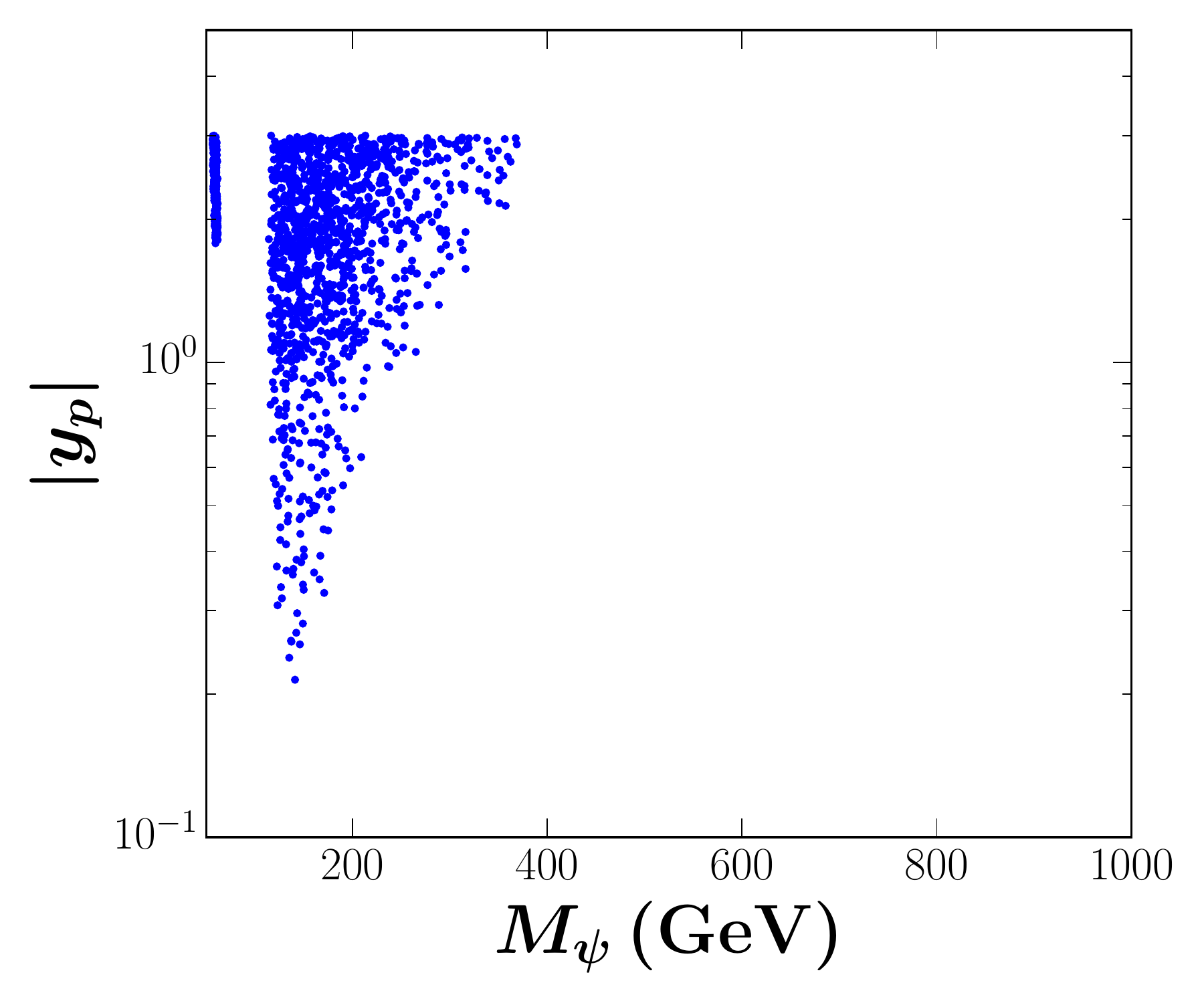}\\
\includegraphics[scale=0.4]{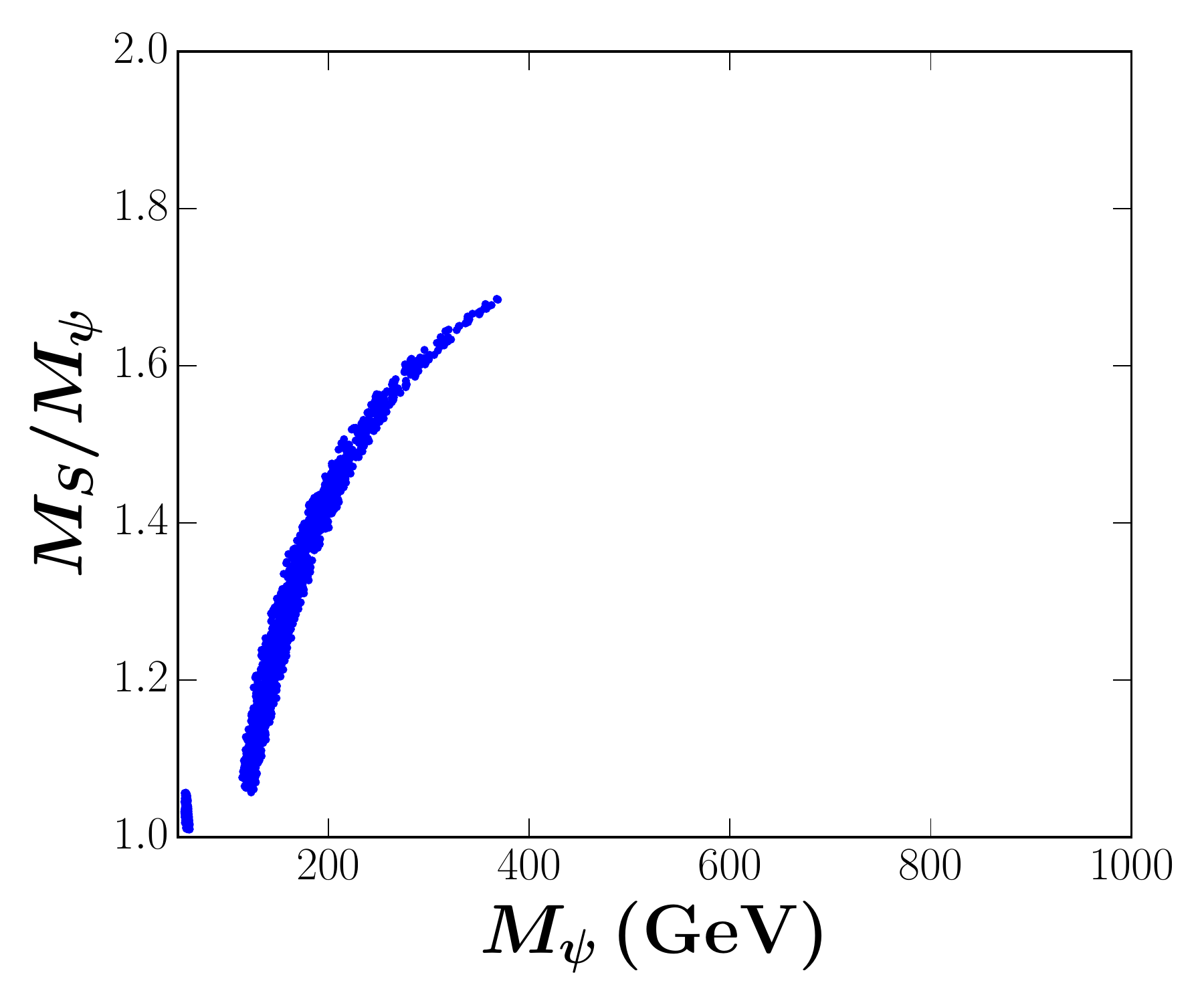}\\
\includegraphics[scale=0.4]{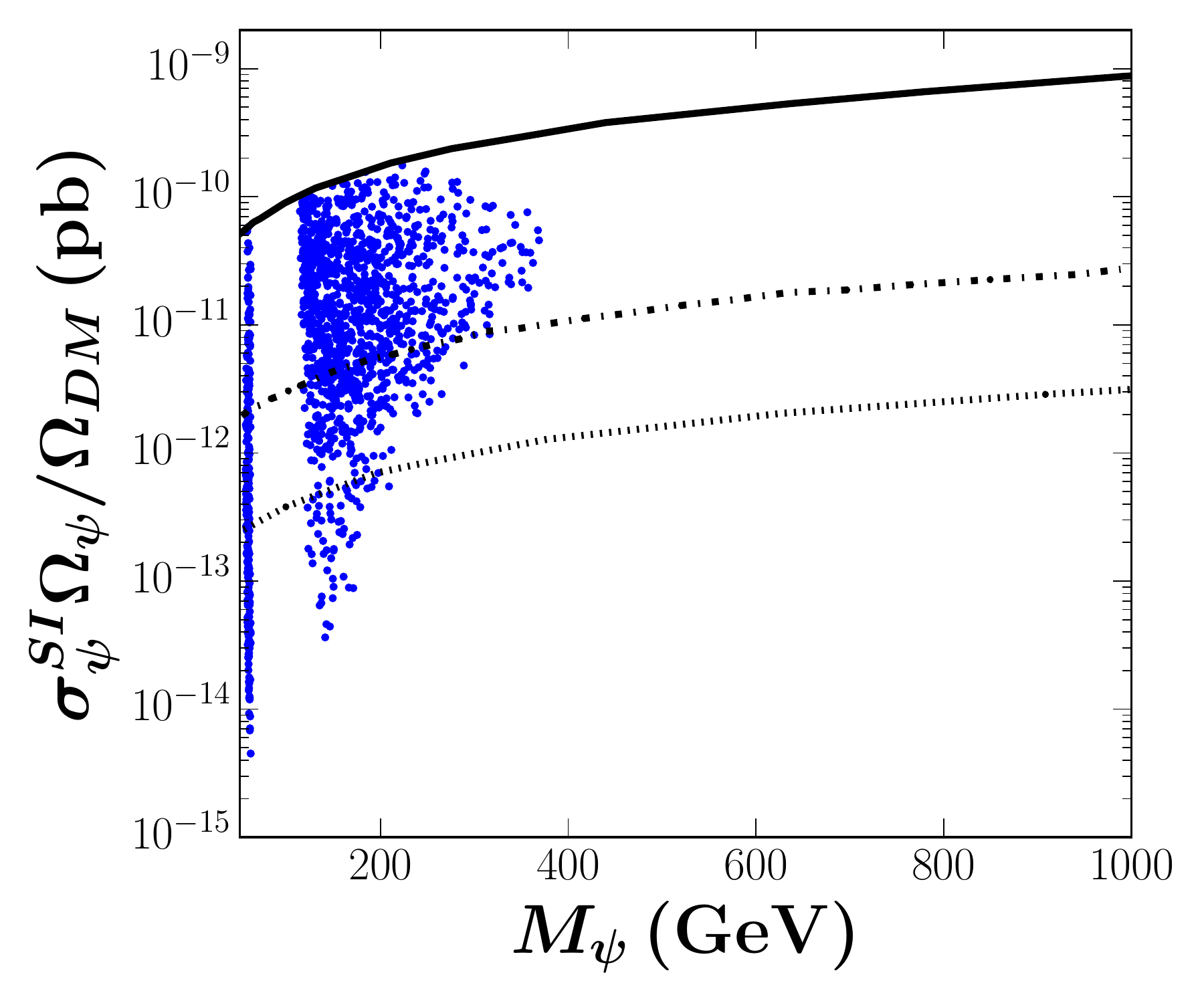}
\includegraphics[scale=0.4]{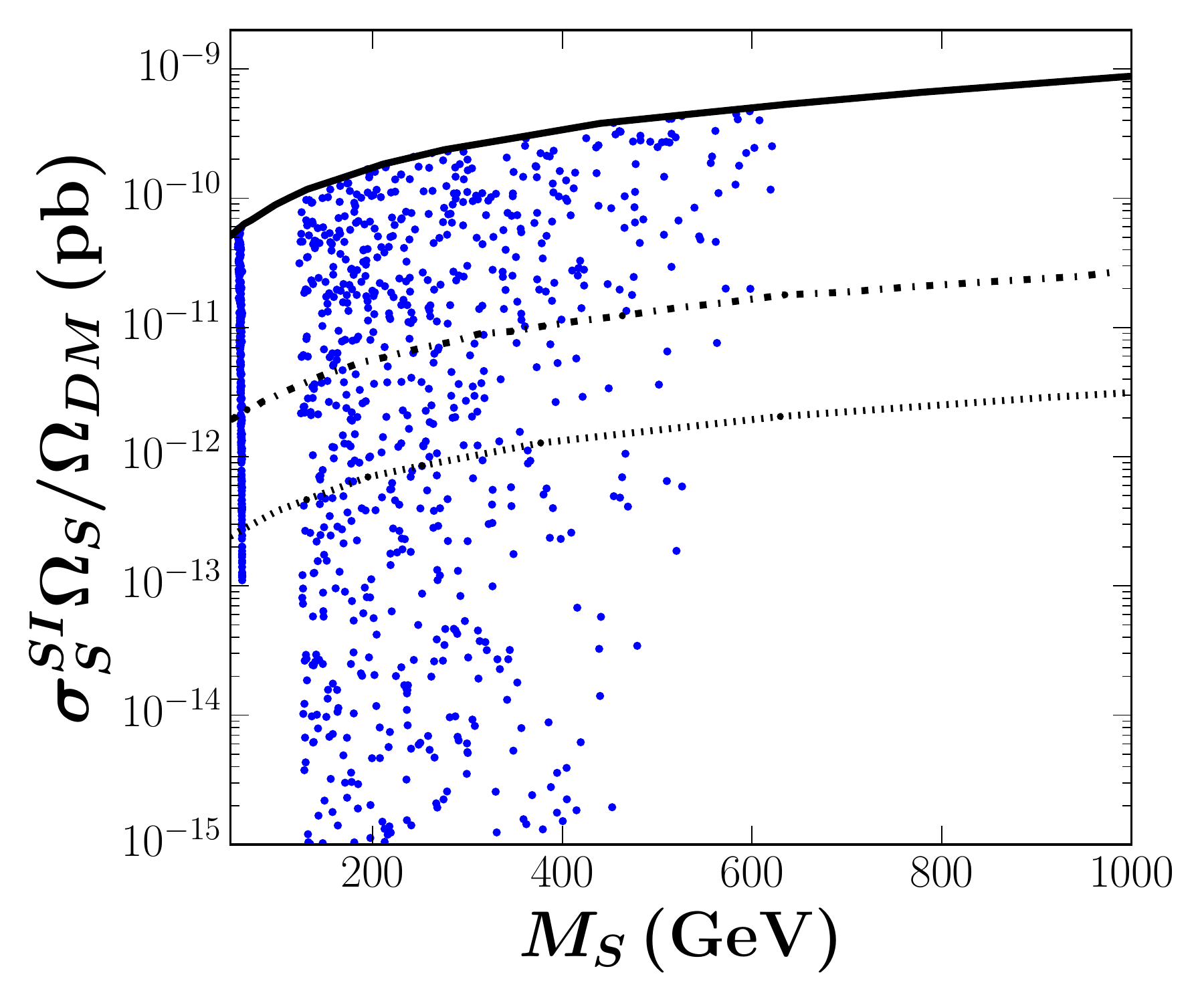}
\caption{A sample of viable models for $M_\psi<M_S$ and $y_s=0$ (pseudoscalar portal) projected onto different planes. The different panels show the couplings $\lambda_{SH}$ (top-left) and $y_s$ (top-right), the ratio of dark matter masses (center), and the direct detection prospects of $\psi$ (bottom-left) and $S$ (bottom-right). In the bottom panels, the lines correspond, from top to bottom, to the current limit from XENON1T, and the expected sensitivities of LZ and DARWIN.}  
\label{fig:ysnull}
\end{figure}

When  $y_s=0$ the only velocity-suppressed process is $\bar \psi \psi \to SS$.  Figure \ref{fig:ysnull} is analogous to figure \ref{fig:ypnull}, displaying a sample of viable models for $y_s=0$. In this case the viable mass range extends only up to $M_\psi\sim 400$ GeV and $M_S\sim 600$ GeV, due to the stronger direct detection limits. We have checked, in fact, that the relic density constraint can be satisfied over a much wider range of dark matter masses.  The crucial point is that the relic density of the scalar can be significantly reduced only for intermediate fermion dark matter masses, i.e. $100\lesssim M_\psi\lesssim 400$ GeV.  Consequently, the re-scaled spin-independent cross-section of the scalar only becomes consistent with Xenon1T data within such a range. From the center panel, we see that $M_S/M_\psi$ can now take values as high as $1.7$, but it varies only along a narrow band.  Regarding the detection prospects, the bottom panels show that the fermion (left) continues to have excellent prospects of being observed in future experiments, with only few points lying below the sensitivity of DARWIN; for the scalar (right), instead, a non-negligible fraction may escape detection.  At the same time, however, several models feature a scalar cross section right below the current limit so that they could be easily probed with new data.

\subsection{General scalar portal}
\label{sub:general}
\begin{figure}[t!]
\centering
\includegraphics[scale=0.4]{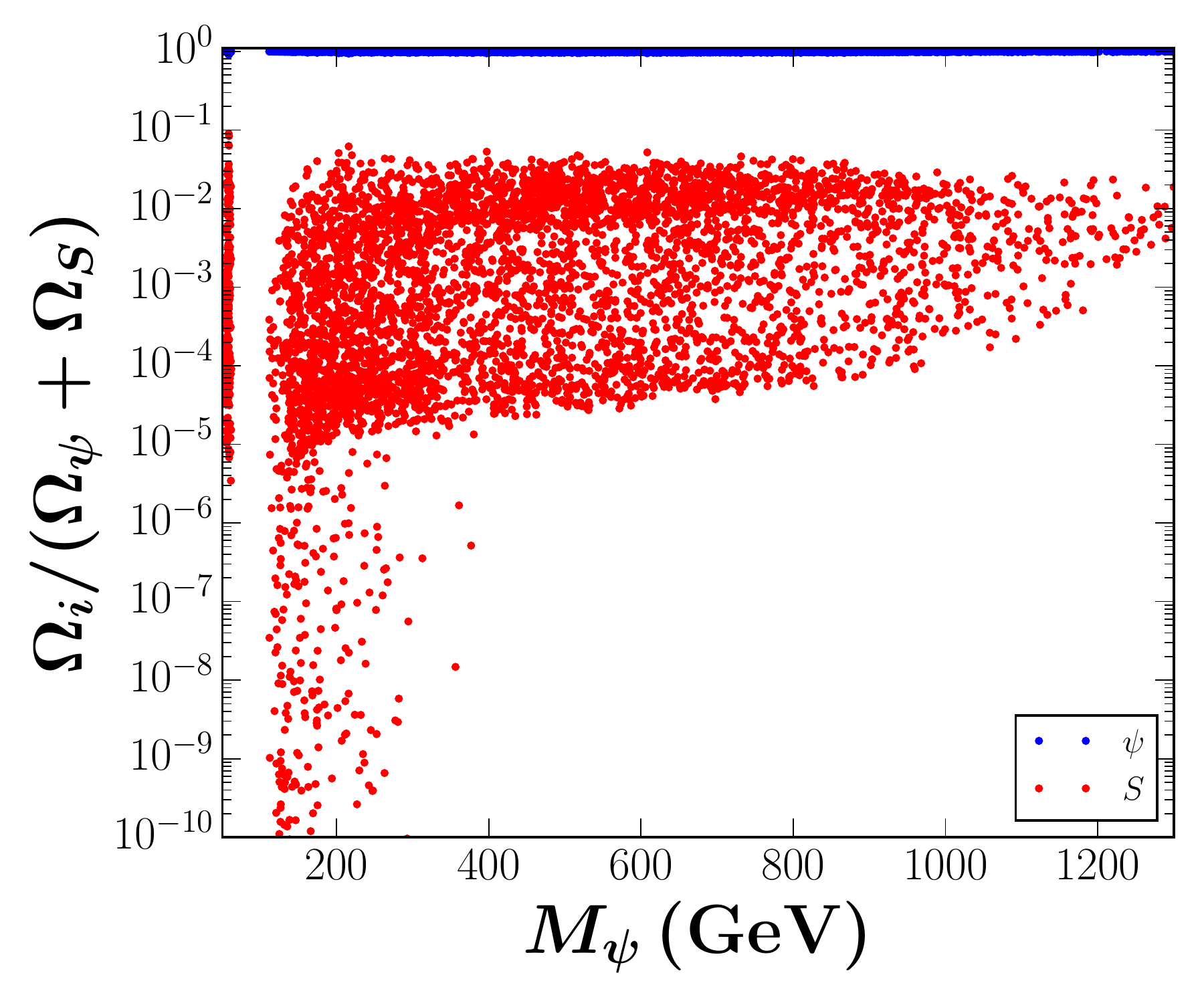}\hspace{0.4cm}
\includegraphics[scale=0.4]{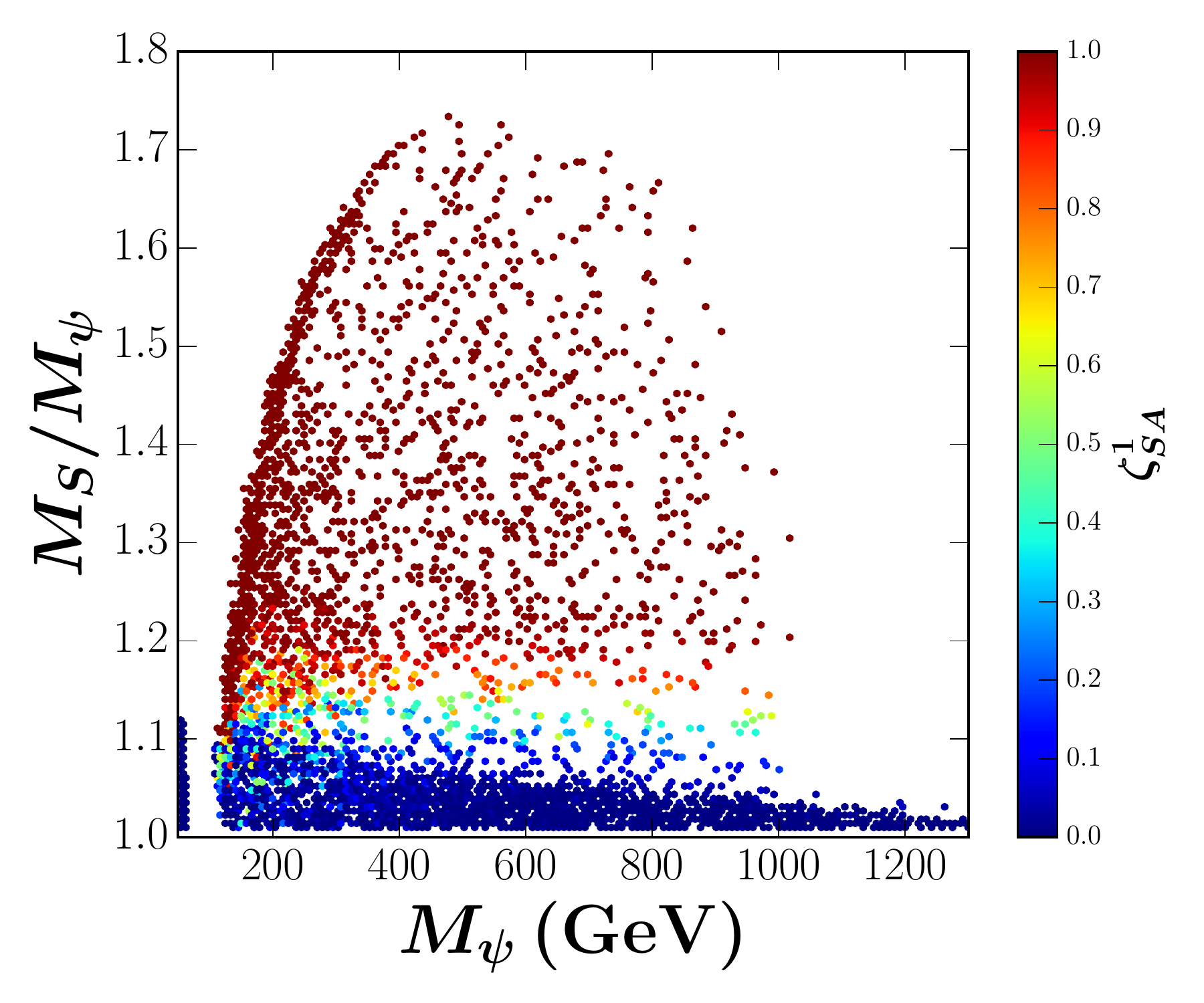}\\
\includegraphics[scale=0.4]{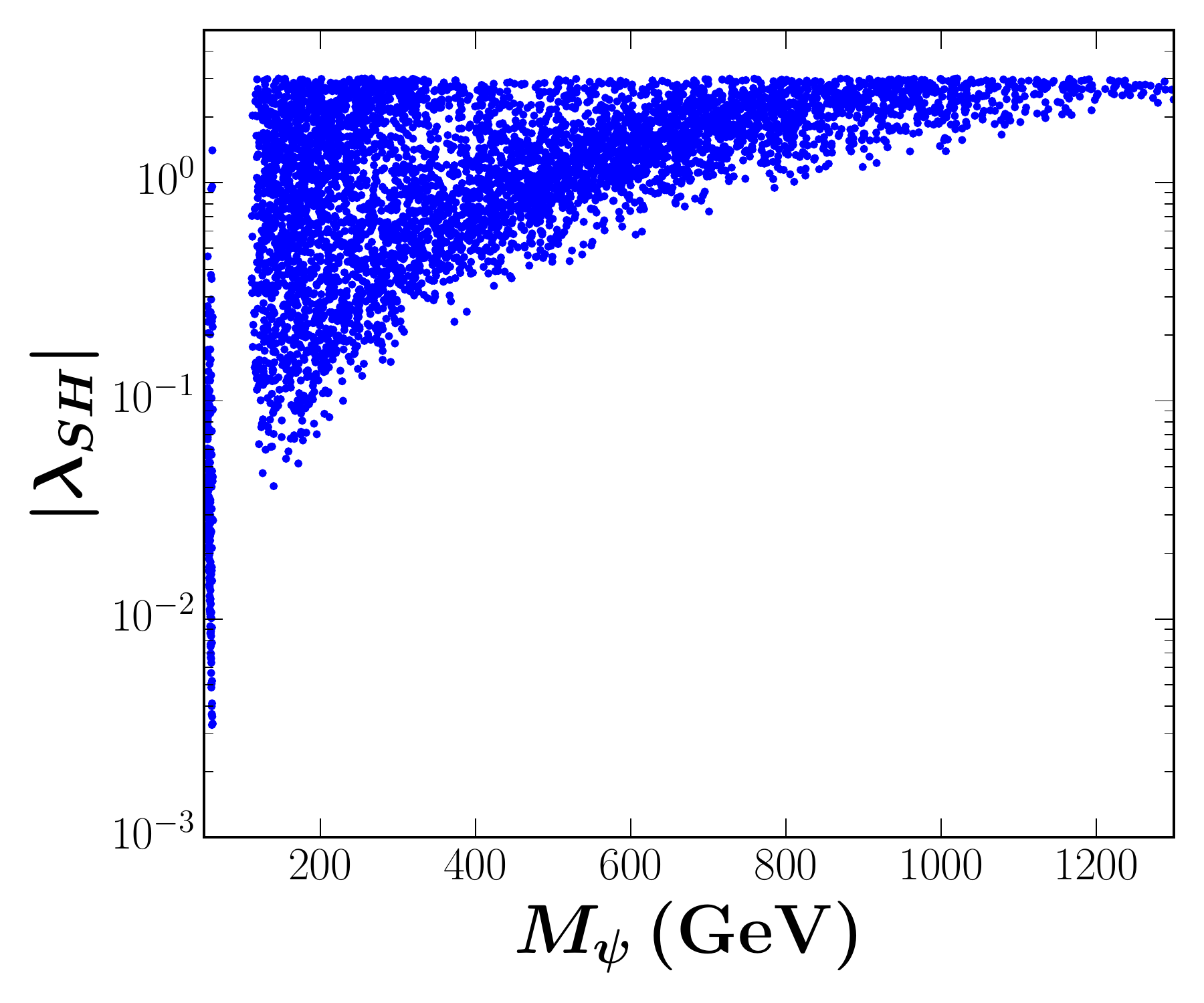}\hspace{0.4cm}
\includegraphics[scale=0.4]{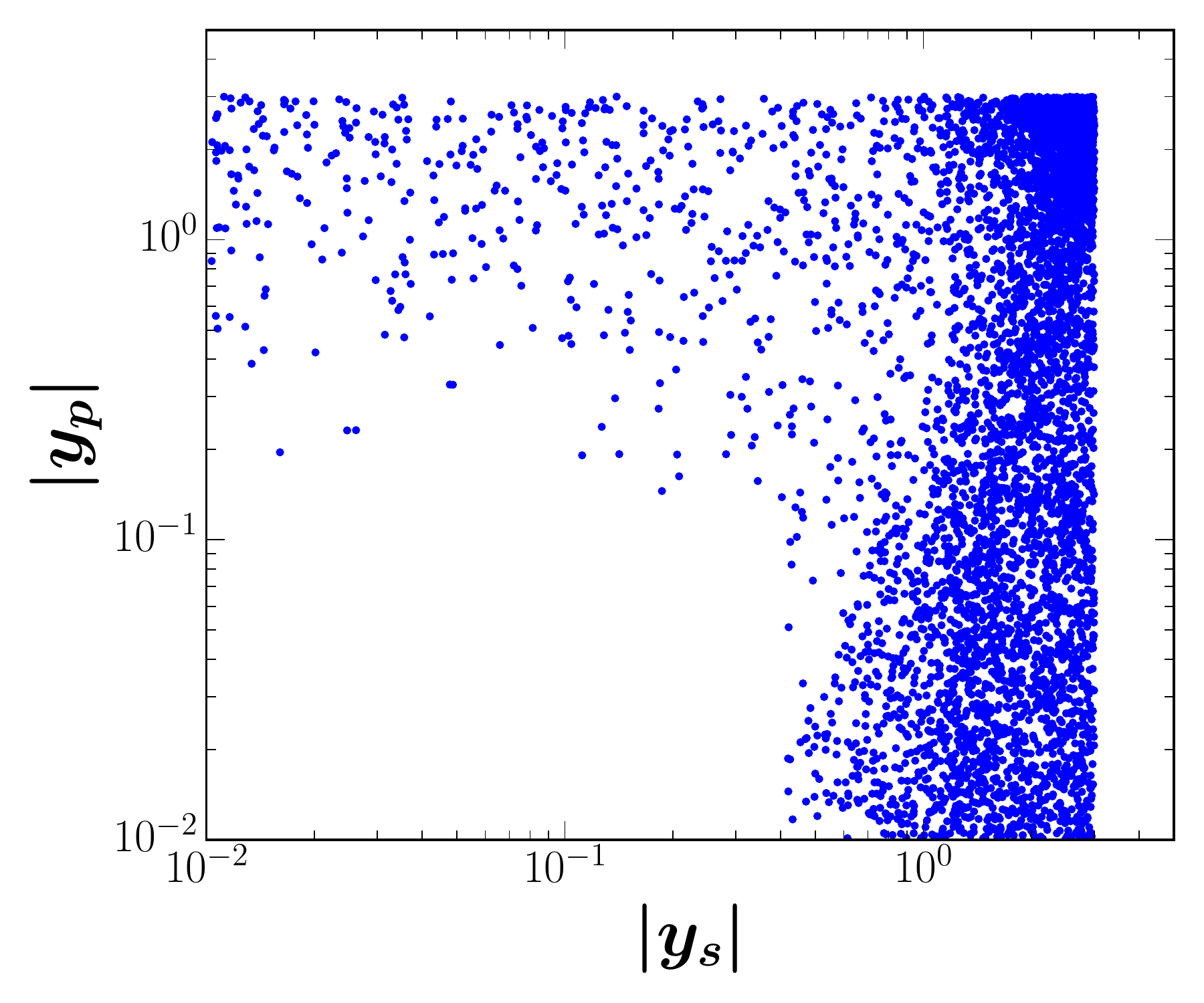}\\
\includegraphics[scale=0.4]{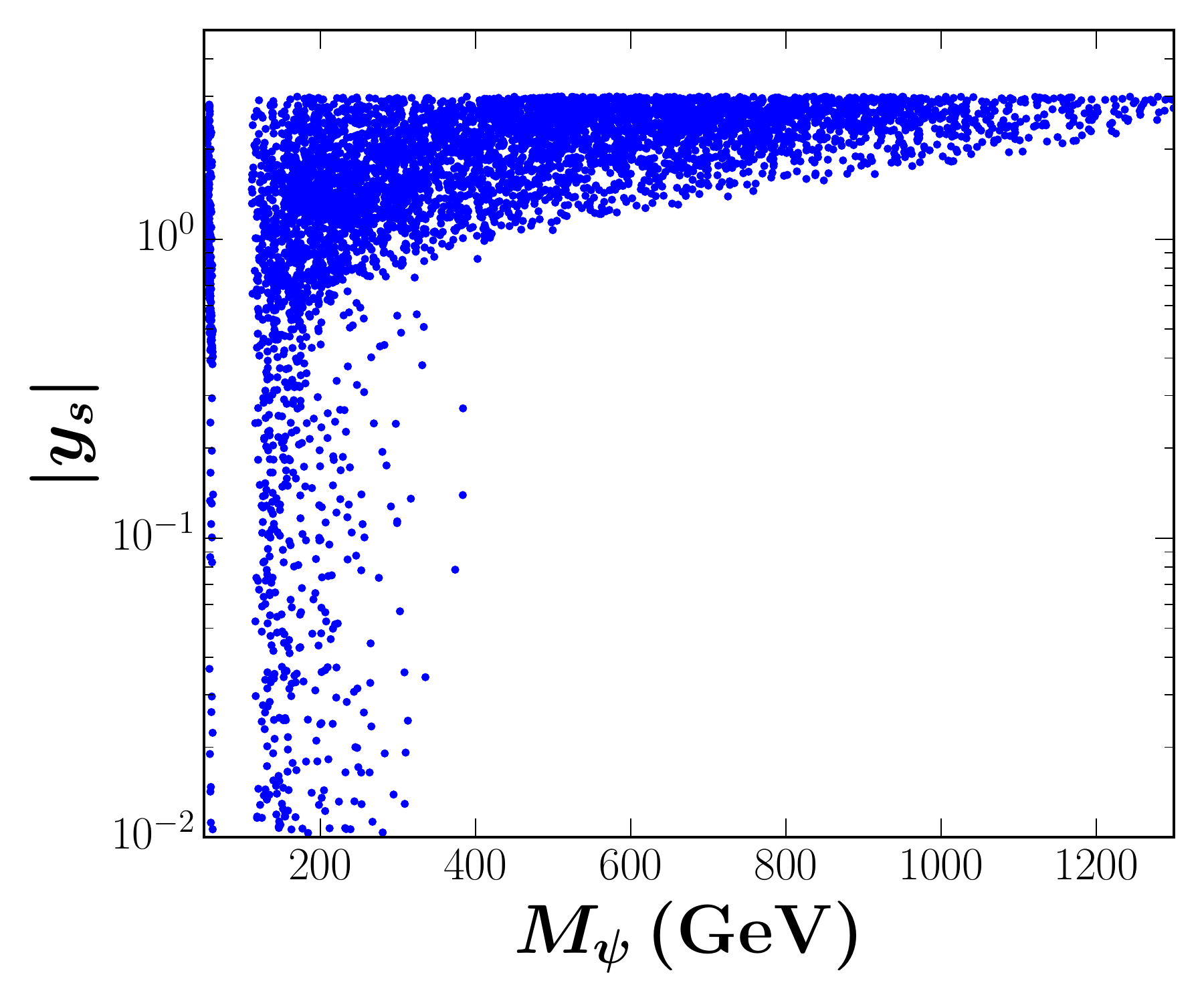}\hspace{0.4cm}
\includegraphics[scale=0.4]{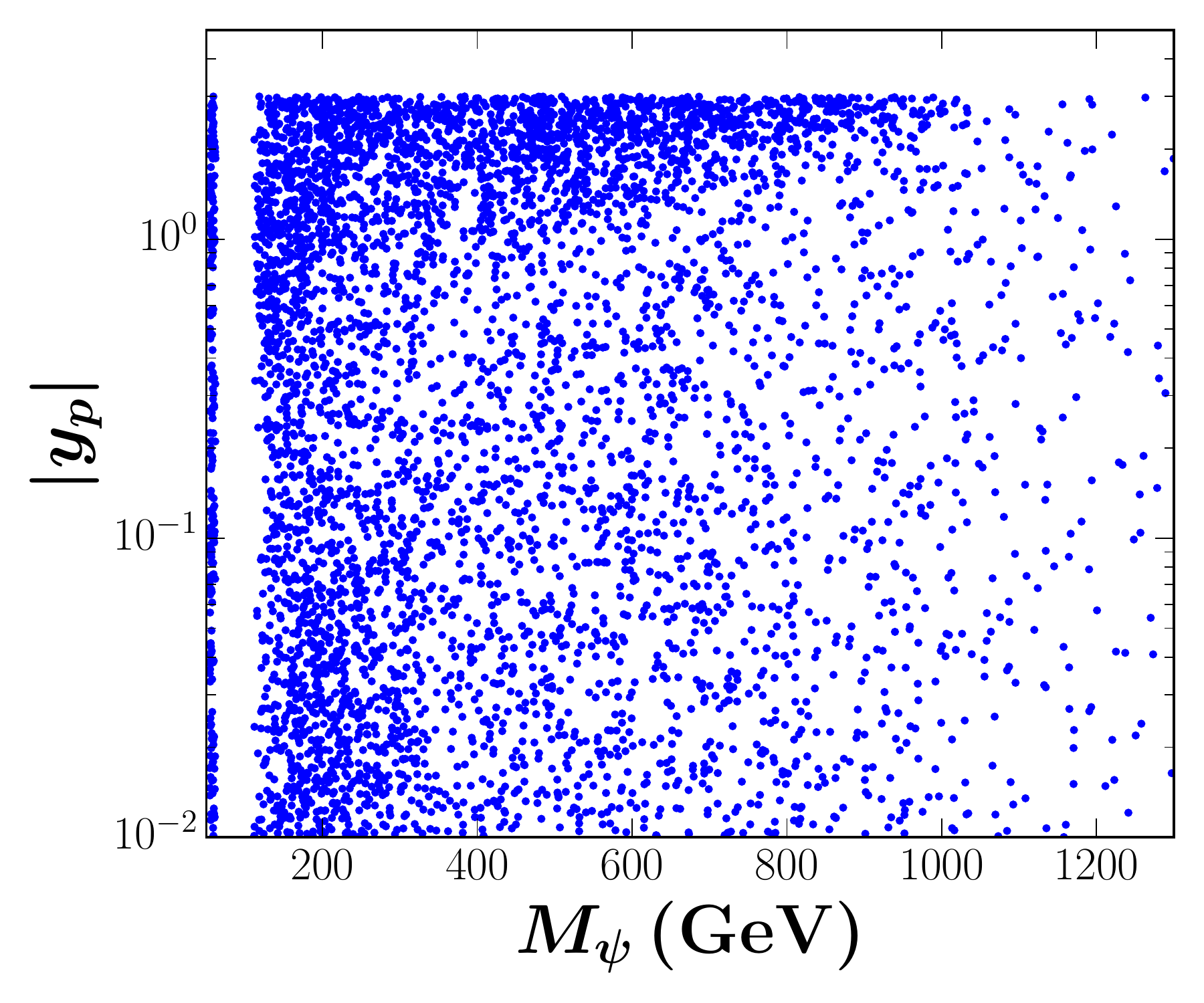}
\caption{A sample of viable models for $M_\psi<M_S$  projected onto different planes. The different panels show the $\psi$ and $S$ contributions to the dark matter density (top-left), the ratio of dark matter masses (top-right), and the different couplings  (center and bottom panels).}  
\label{fig:yyy}
\end{figure}

\begin{figure}[tp]
\centering
\includegraphics[scale=0.4]{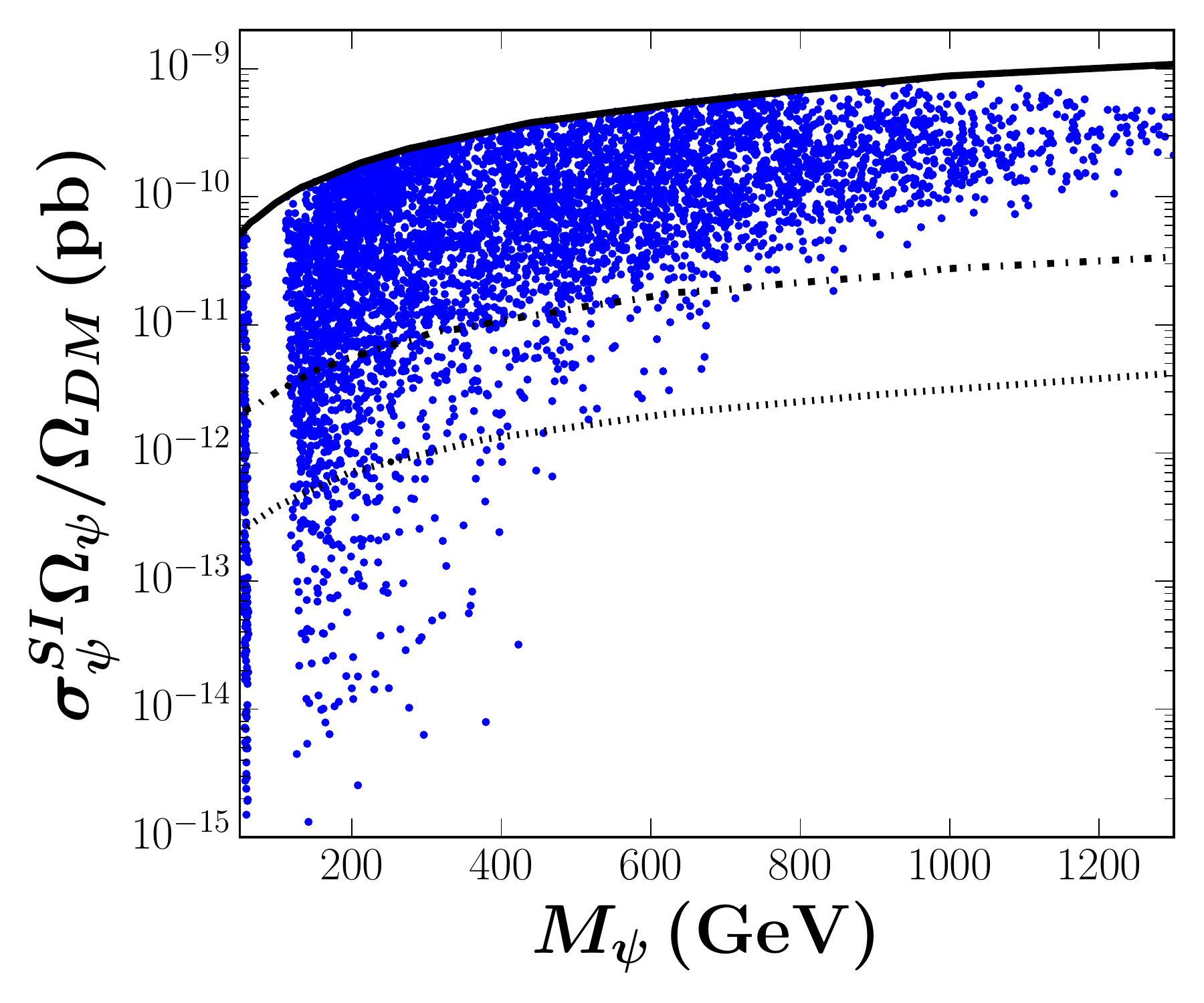}
\includegraphics[scale=0.4]{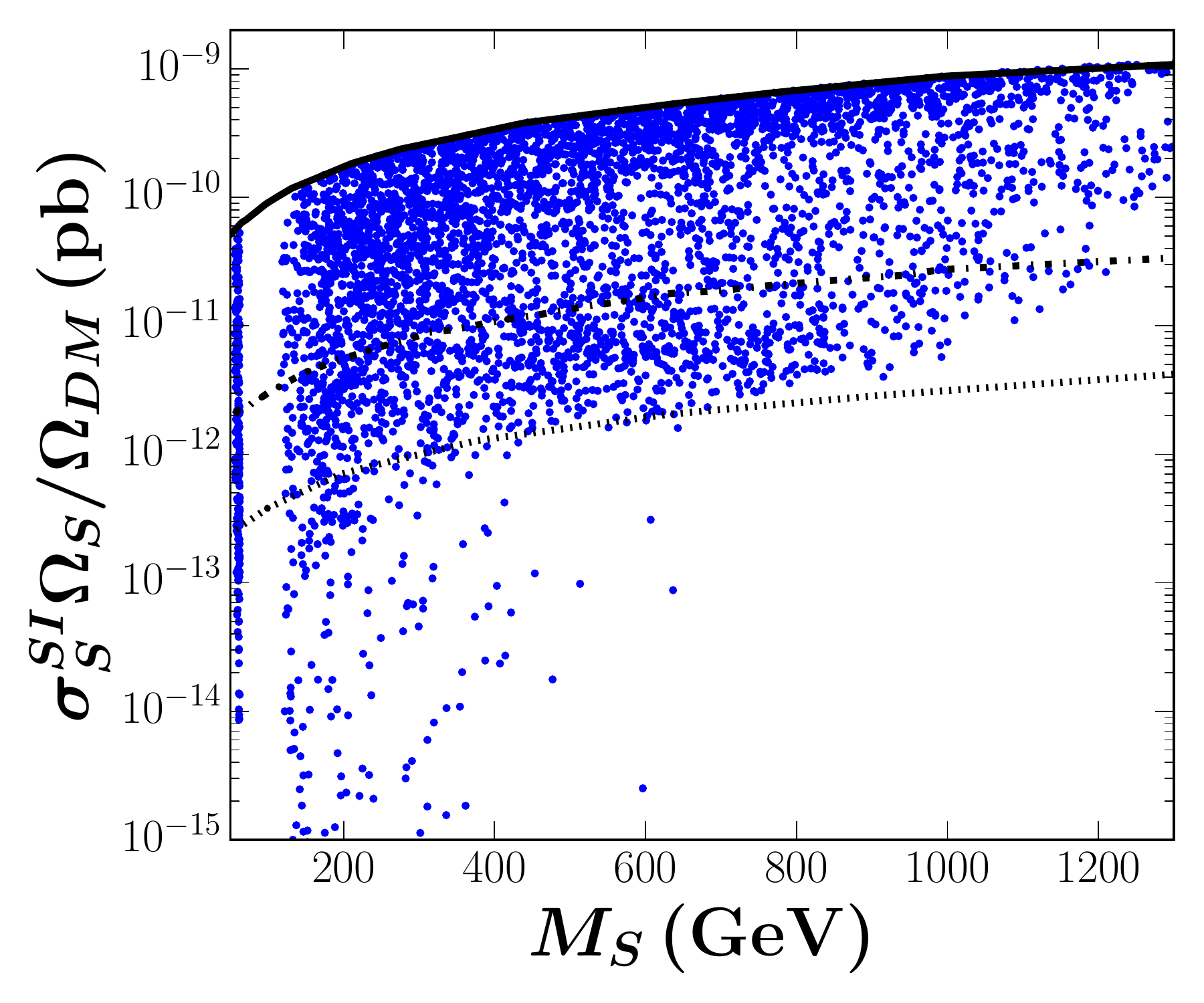}\\
\includegraphics[scale=0.4]{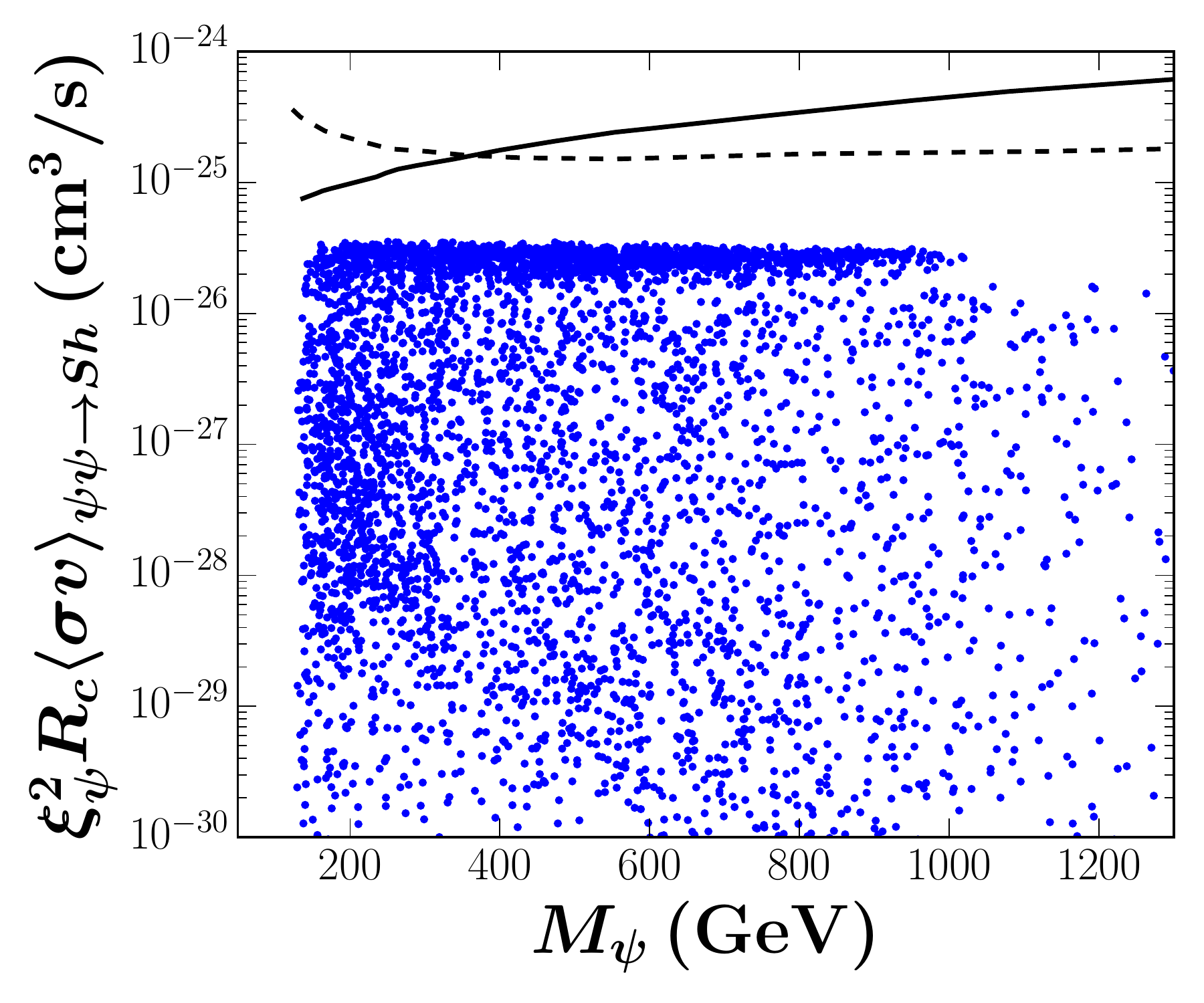}
\caption{The direct (top) and indirect (bottom) dark matter detection rates for our sample of models in the regime  $M_{\psi} < M_{S}$. In the top panels, the lines correspond, from top to bottom, to the current limit from XENON1T, and the expected sensitivities of LZ and DARWIN. In the bottom panel, the lines indicate the Fermi-LAT limit (solid) and the expected sensistivy of CTA (dashed).}  
\label{fig:ID}
\end{figure}

Let us now consider the general case for the $M_\psi<M_S$ regime. Figures \ref{fig:yyy} and \ref{fig:ID} show a sample of viable models projected onto different planes. The top-right panel of figure \ref{fig:yyy} displays, as a function of $M_\psi$, the  fraction of the dark matter density that is due to $\psi$ (blue) and $S$ (red). We see that the scalar always contributes less than $10\%$ of the dark matter density, with most points concentrated between fractions of order $10^{-2}$ and $10^{-5}$ over the entire range of $M_\psi$. For $M_\psi\lesssim 400$ GeV, the scalar contribution can be significantly smaller, reaching values as low as $10^{-10}$. It is clear then that it is the fermion that accounts for most of the dark matter. Let us stress, though, that this does not imply that the scalar can be neglected because, as will be shown, it can lead to observable signals in dark matter experiments. 

Semiannihilations play a vital role in this model as they allow the fermion relic density to decrease significantly --a fact that was recognized already in \cite{Cai:2015zza}. To quantify their relevance, it is useful to define the semiannihilation fractions for the two dark matter particles as  
\begin{align}
    \zeta_{SA}^1&\equiv\frac{\frac{1}{2}\sigma_v^{1120}}{\frac{1}{2}\sigma_v^{1120}+\sigma_v^{1122}},\quad
    \zeta_{SA}^2\equiv\frac{\frac{1}{2}\sigma_v^{1210}}{\sigma_v^{2200}+\frac{1}{2}\sigma_v^{1210}+\sigma_v^{2211}}.
\end{align}
The top-right panel of figure \ref{fig:yyy} shows the ratio $M_S/M_\psi$, with the value of $\zeta_{SA}^1$ color-coded according to the scale on the right. Unlike for the scalar and pseudoscalar portals,  in this case $M_S/M_\psi$ can reach sizable values (of order $1.7$) up to $M_\psi\sim 1$ TeV. Over that range and for $M_S/M_\psi\gtrsim 1.1$  semiannihilations are seen to be the dominant mechanism responsible for the $\psi$ relic density.  It is only between $1$ and $1.3$ TeV that dark matter conversions becomes dominant and that $S$ and $\psi$ are required to be highly degenerate. The difference between this figure and the analogous one for the scalar portal, figure \ref{fig:ypnull}, demonstrates that $y_p$, which was not considered in \cite{Cai:2015zza}, plays a non-negligible role in the phenomenology of the model. Indeed, there exists viable regions of the parameter space that can be reached only if $y_p\neq 0$. 

The allowed values for the couplings in our sample of viable models are illustrated in the center and bottom panels of figure \ref{fig:yyy}.  As expected, either $y_s$ or $y_p$ must be sizable ($\gtrsim 0.1$), along with $\lambda_{SH}$. The highest  $\psi$  mass in our sample corresponds to the region where $\lambda_{SH}$ and $y_s$ both reach the maximum value permitted by our scan.

Figure \ref{fig:ID} shows the direct and indirect detection prospects within our sample of viable models. The top panels compare, for $\psi$ (left) and $S$ (right), the elastic scattering cross section against the current limit set by XENON1T (solid line) and the expected sensitivities in LZ and DARWIN. From the figures we see that most models in our sample lie within the expected sensitivity of DARWIN and that a significant fraction of them feature cross sections just below the current limit. This regime, therefore, offers excellent prospects to be tested in current and planned direct detection experiments. And for dark matter masses below $1$ TeV it may be possible, thanks to the mass difference, to distinguish the signal produced by each dark matter particle, excluding in this way the standard scenario with one dark matter particle.

Regarding indirect detection, the most promising process in both mass regimes is $\psi\psi\to Sh$. 
Up to date there is no reported experimental limit on such a process, however it is possible to set a the limit from the related process $DM+DM\to DM+h$ \cite{Queiroz:2019acr}  by rescaling \cite{DiazSaez:2021pmg} 
 the cross section $\langle\sigma v\rangle_{\psi\psi\to Sh}$ by the factor 
 \begin{align}
     R_c=
     \frac{(M_h^2-M_S^2-4M_\psi^2)^2}{(M_h^2+3M_\psi^2)^2}. 
 \end{align}
 The corresponding results are displayed in the bottom panel of figure \ref{fig:ID} (blue points), with the solid (dashed) blue line being the limit (prospect) extracted in Ref.~\cite{Queiroz:2019acr} using the data reported by the Fermi collaboration \cite{Ackermann:2015zua}  (and the sensitivity of the Cherenkov Telescope Array CTA \cite{Silverwood:2014yza}).   
\section{The $M_S<M_\psi$ regime}\label{sec:pheno2}
\begin{figure}[tp]
\centering
\includegraphics[scale=0.4]{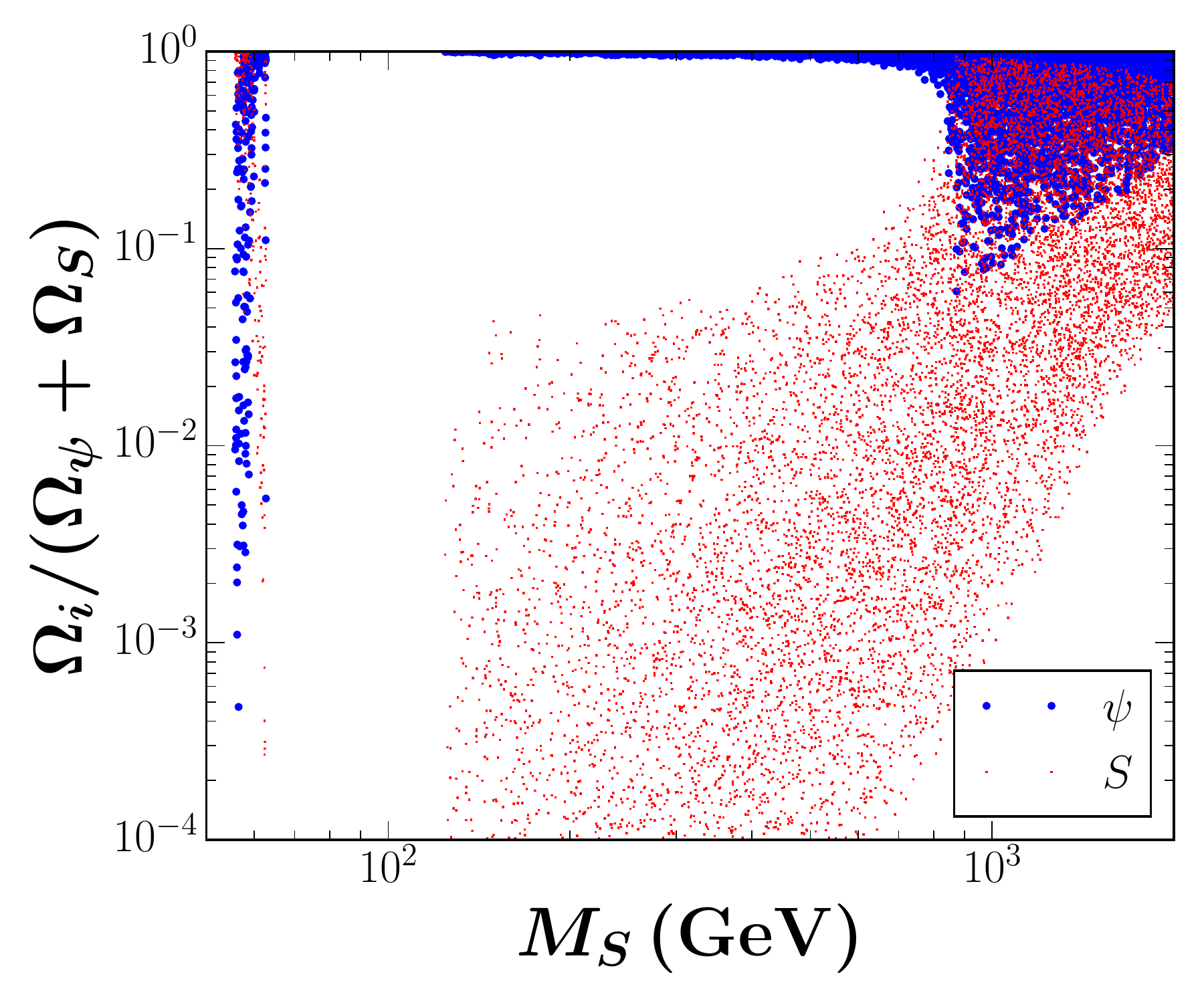}
\includegraphics[scale=0.4]{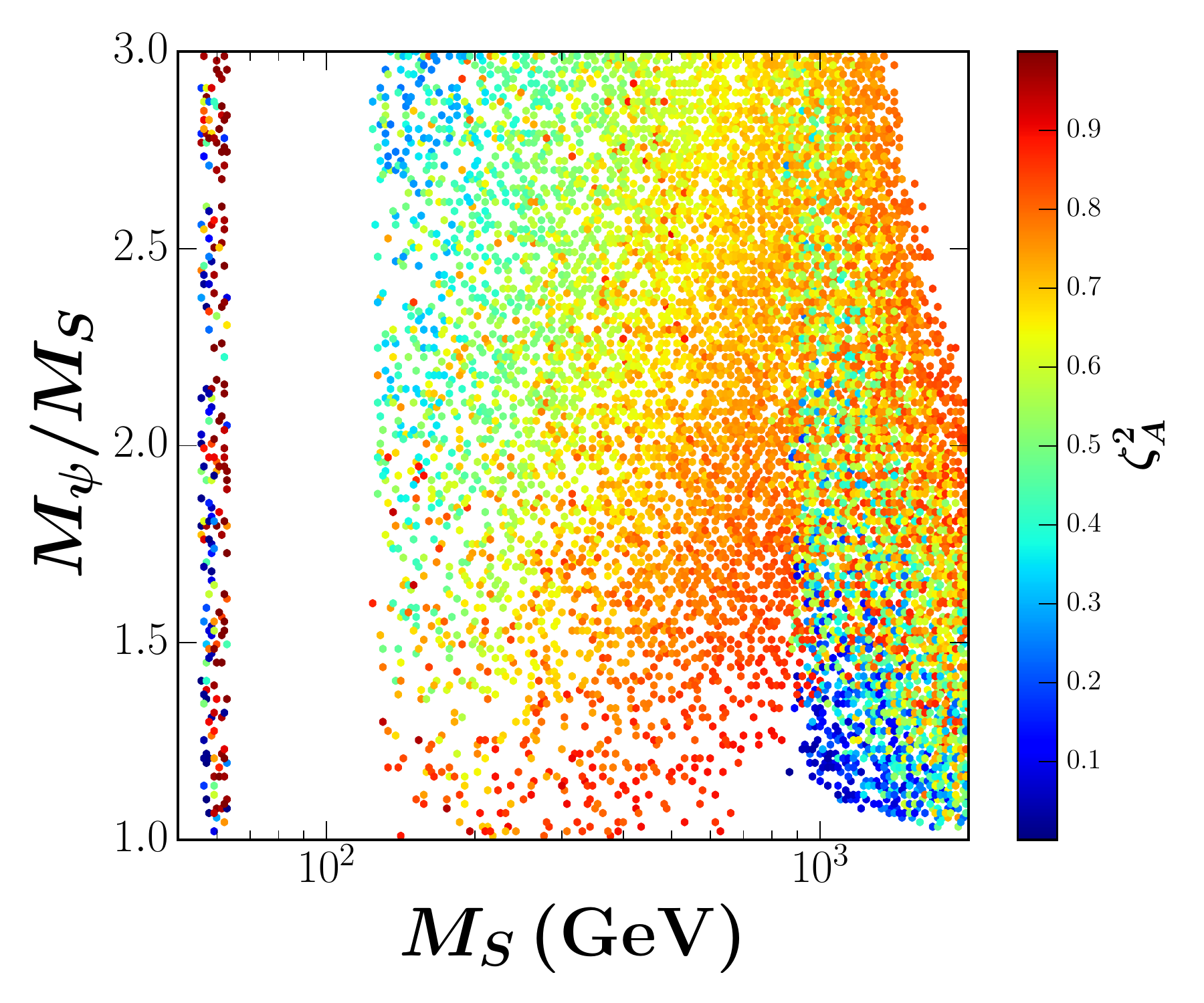}\\
\includegraphics[scale=0.4]{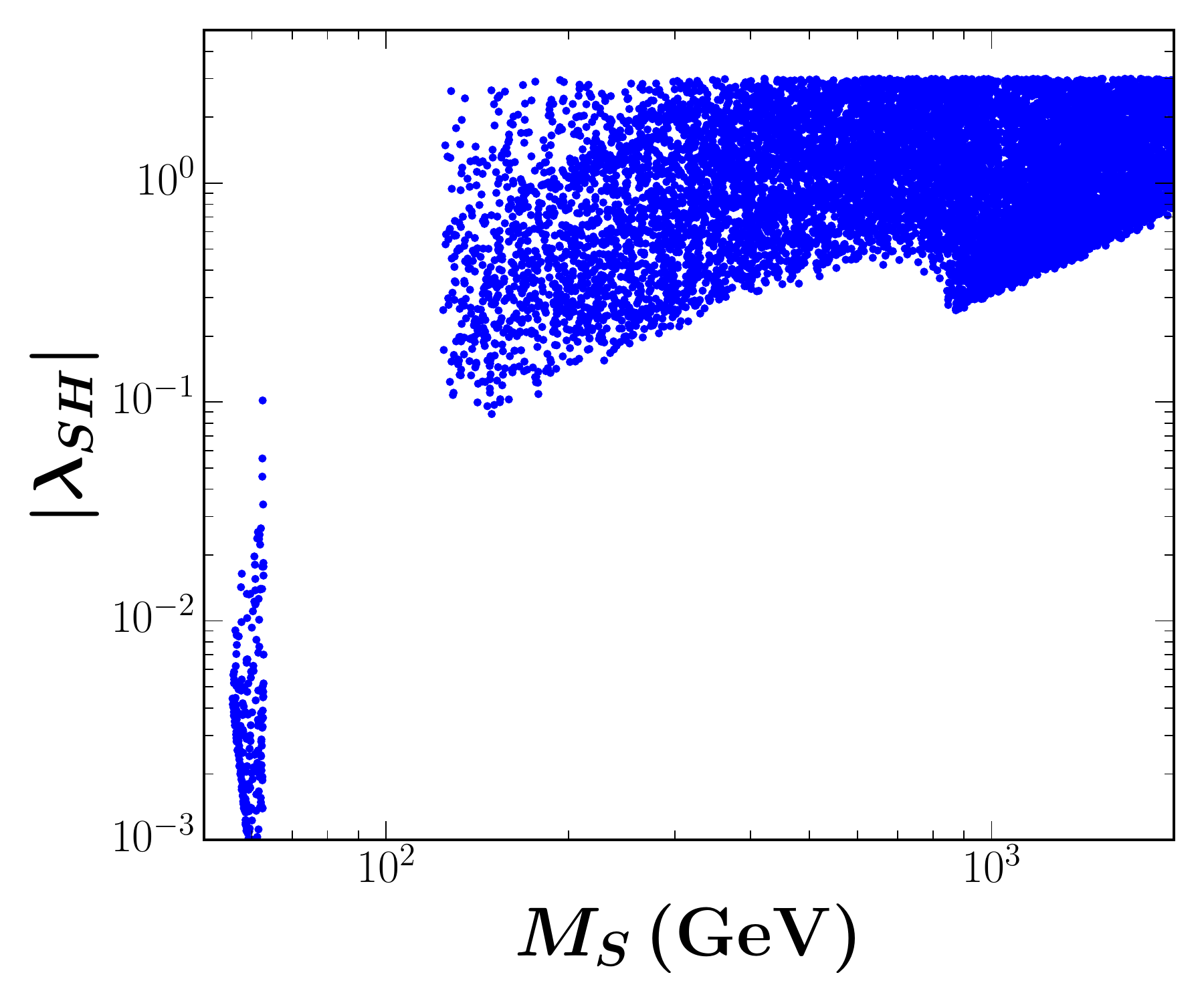}
\includegraphics[scale=0.4]{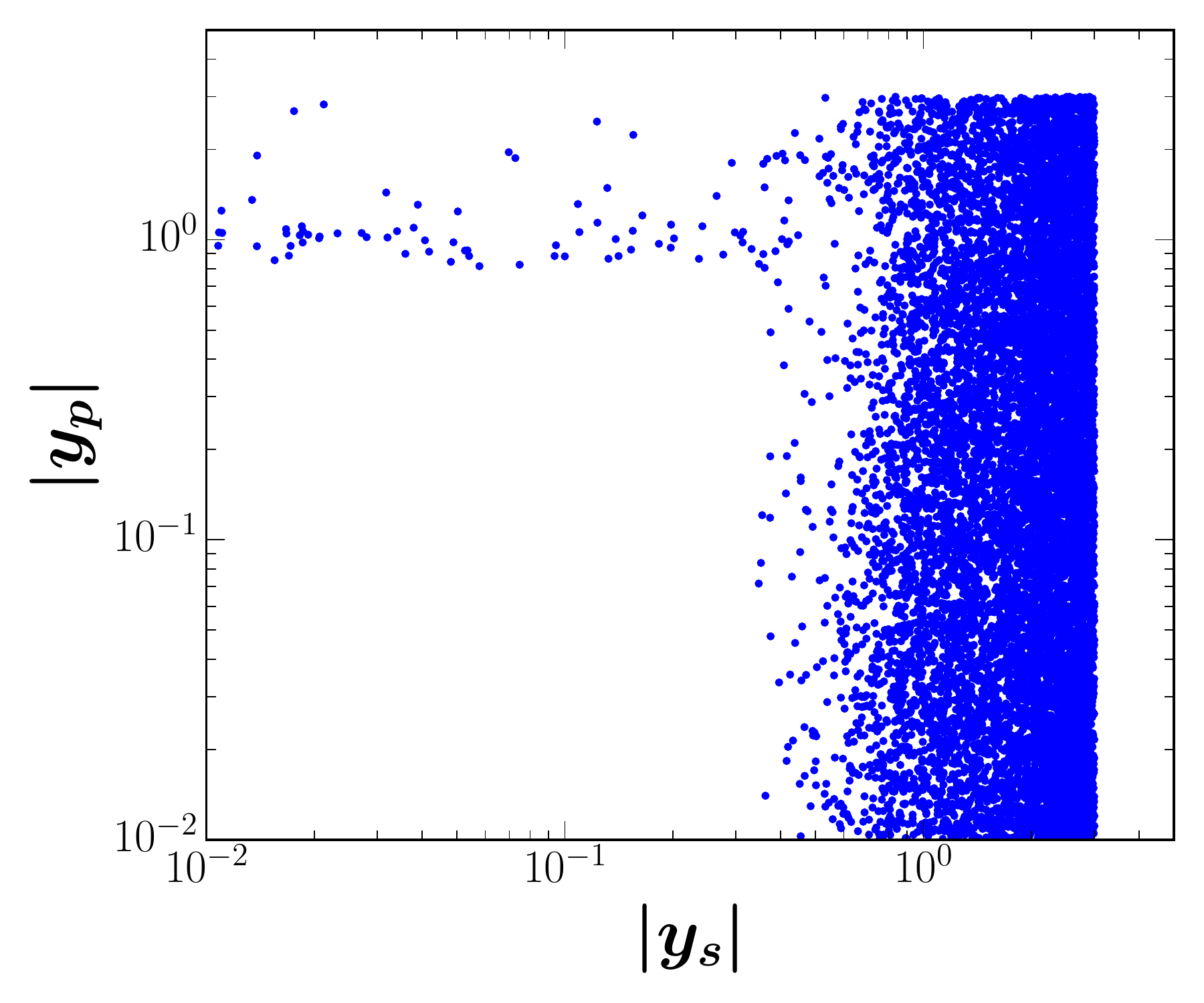}\\
\includegraphics[scale=0.4]{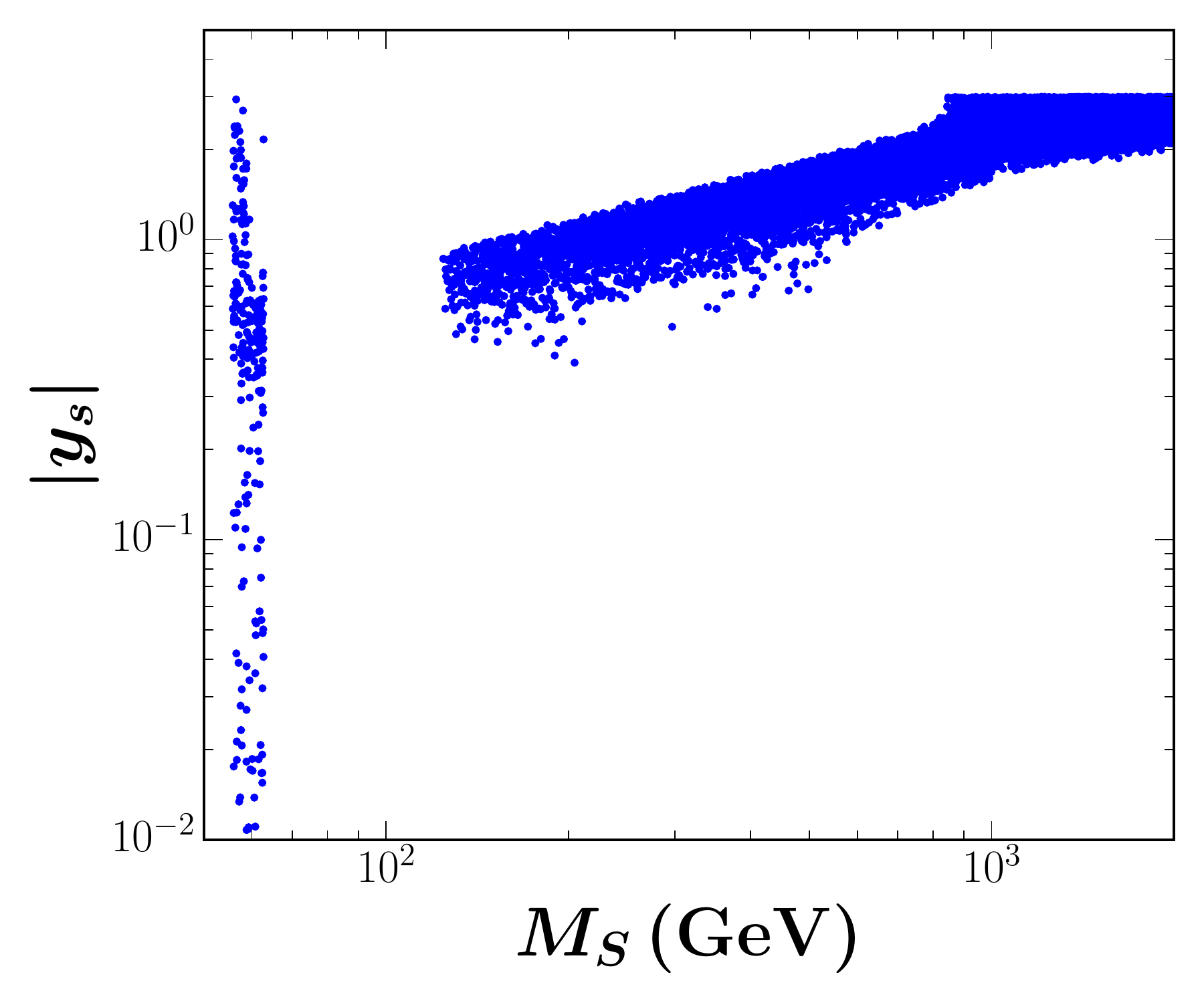}
\includegraphics[scale=0.4]{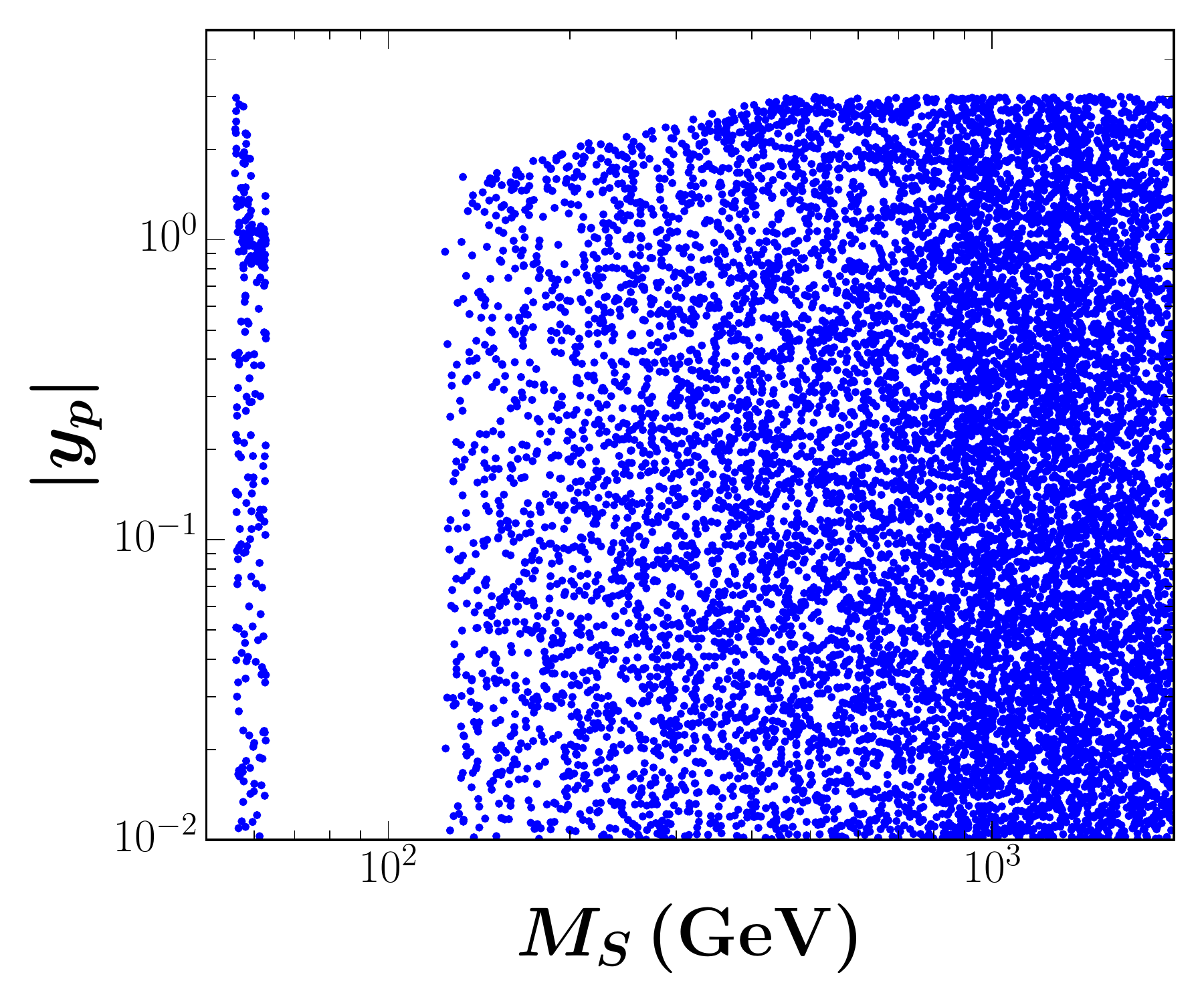}
\caption{A sample of viable models for $M_S<M_\psi$  projected onto different planes. The different panels show the $\psi$ and $S$ contributions to the dark matter density (top-left), the ratio of dark matter masses (top-right) and the allowed range of variation for the three couplings (center and bottom panels). }  
\label{fig:lop-S2}
\end{figure}

\begin{figure}[tp]
\centering
\includegraphics[scale=0.4]{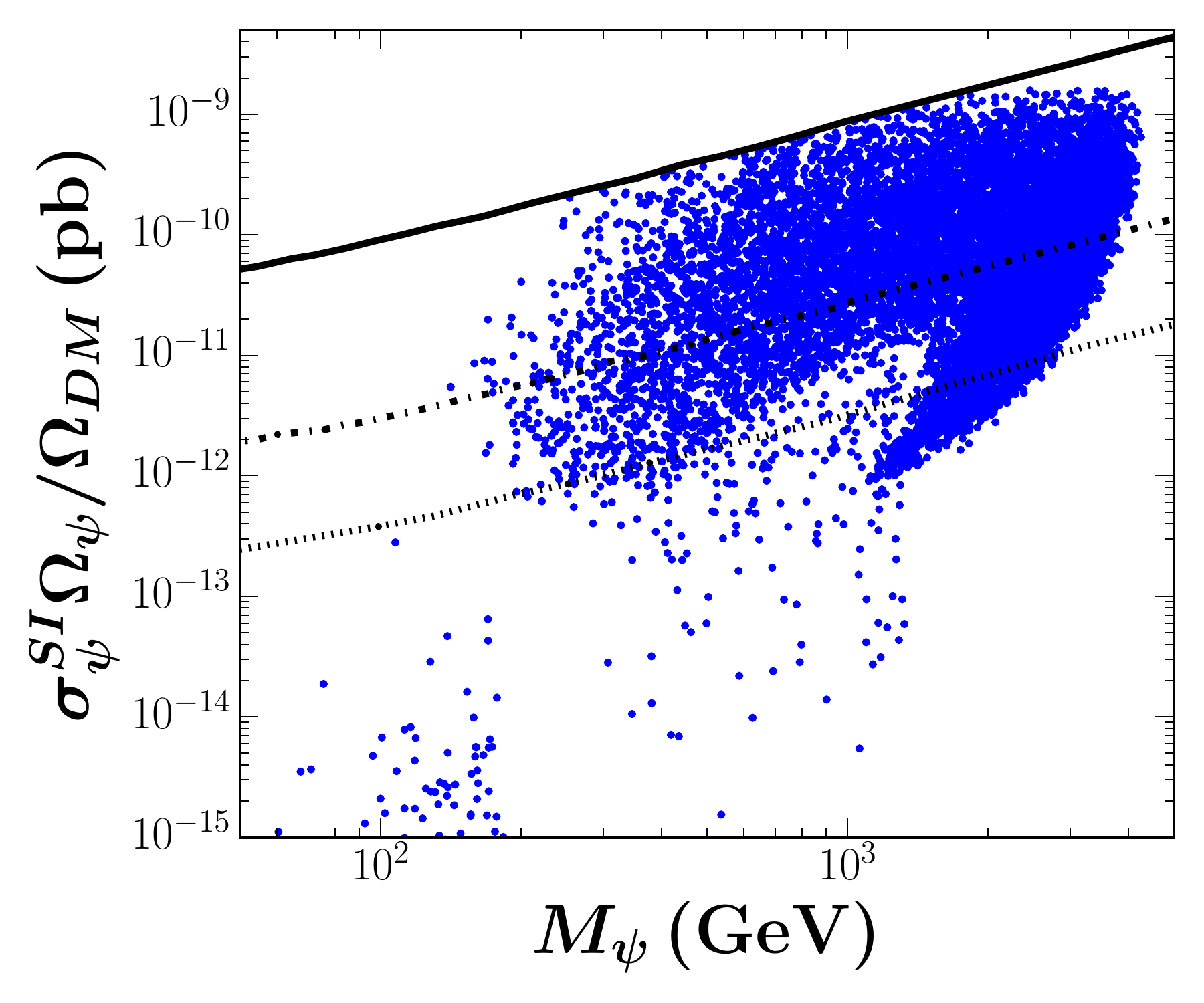}
\includegraphics[scale=0.4]{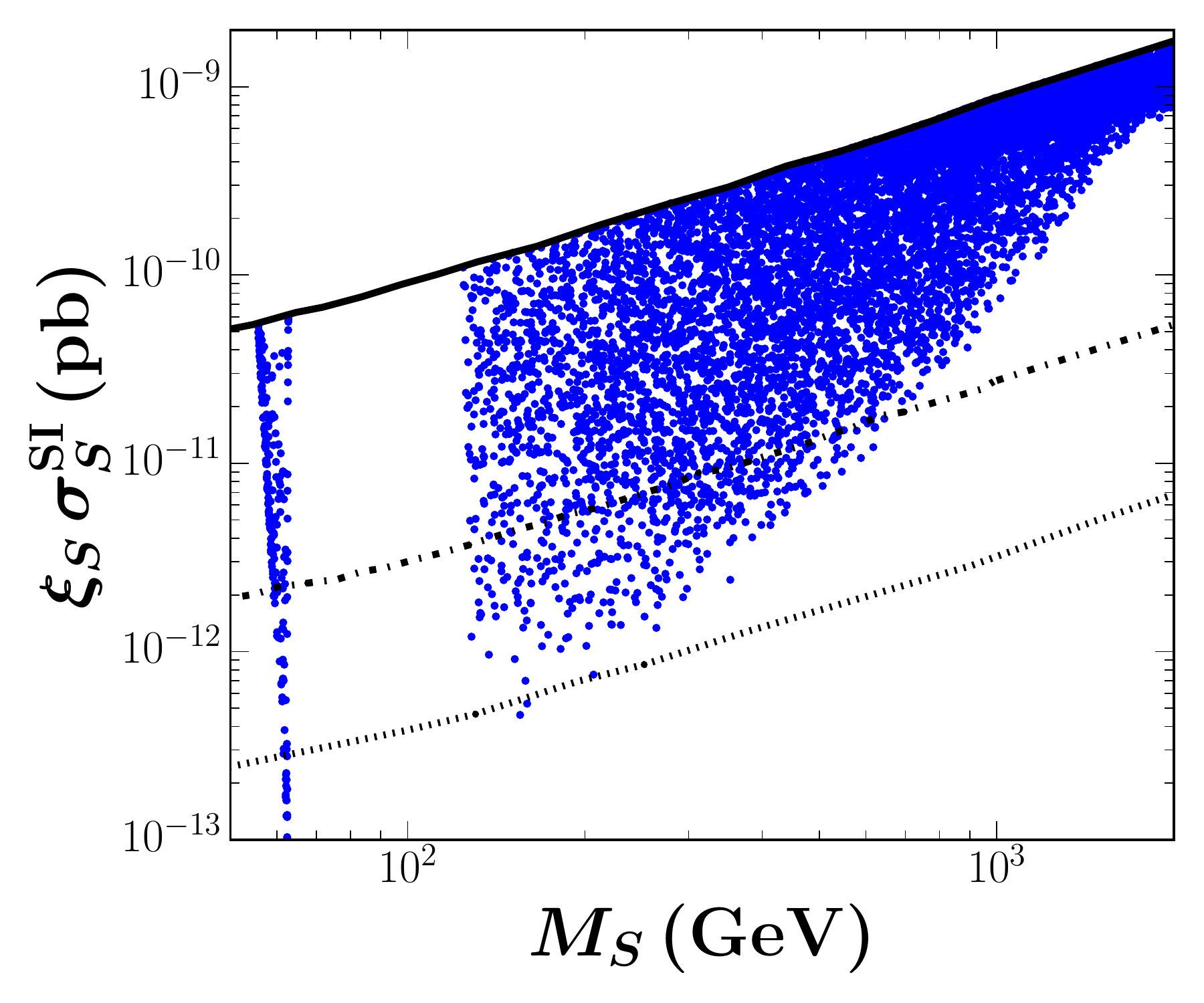}\\
\includegraphics[scale=0.4]{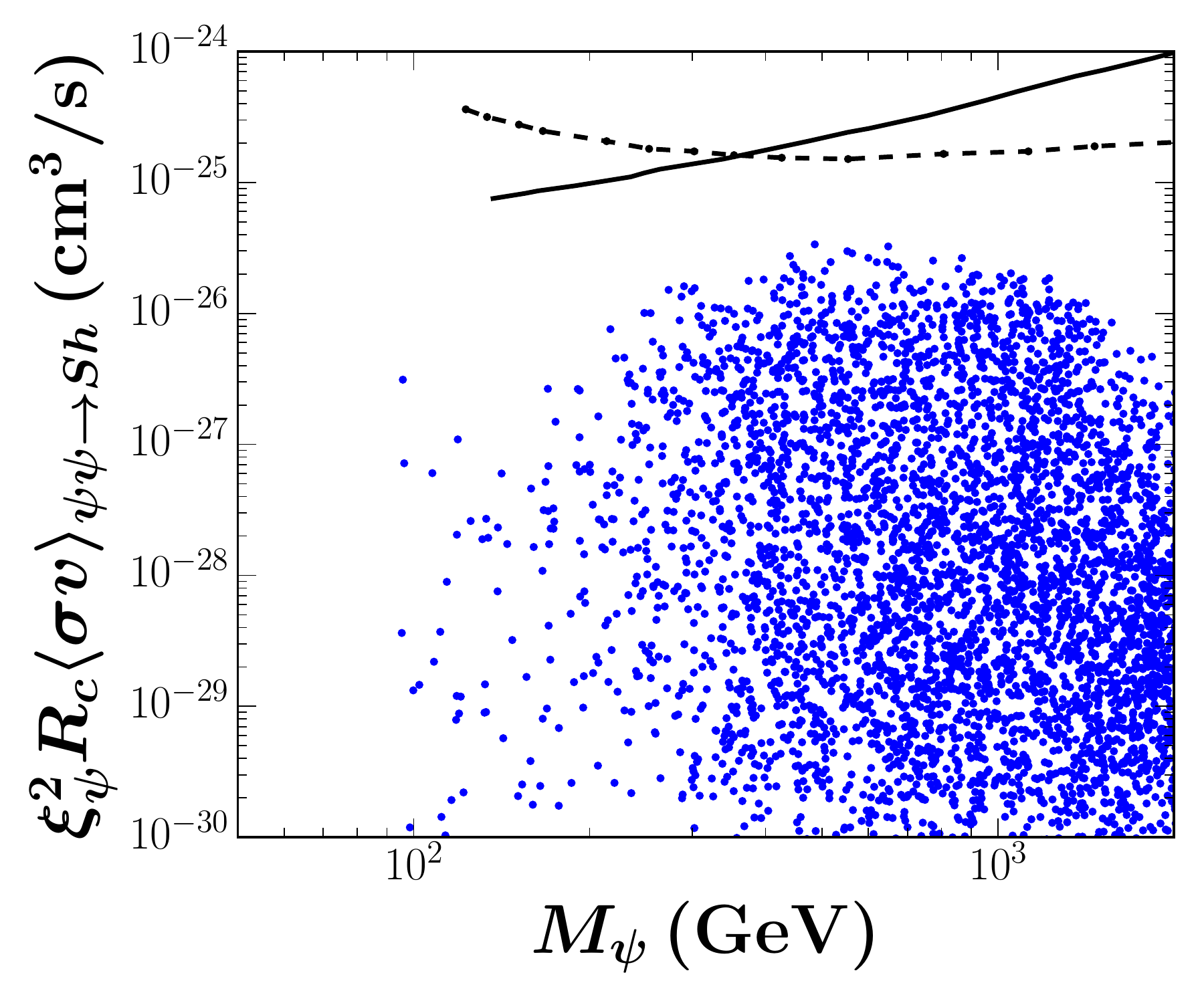}
\caption{The detection prospects in our sample of viable models for $M_S<M_\psi$. The top panels show the direct detection cross sections for the fermion (left) and the scalar (right). From top to bottom, the lines correspond to the current limit from XENON1T and the expected sensitivities of LZ and DARWIN. The bottom panel illustrates the indirect detection signal due to the process $\psi+\psi\to S+h$. The lines correspond to the Fermi-LAT limit (solid) and the expected sensistivy of CTA (dashed).}  
\label{fig:ID-lop-S2}
\end{figure}

For this regime, the parameters are randomly varied using a log-uniform distribution within the ranges
\begin{align}
    &50\,{\rm GeV}\leq M_{S}\leq 2\,{\rm TeV},\,\,\,M_S<3M_\psi,\\
    &10^{-3}\leq |\lambda_{SH}|\leq 3,\\
    &10^{-2}\leq |y_s|,|y_p|\leq 3.
\end{align}
Because the pseudoscalar portal ($y_s=0$) turned out to be viable only at the Higgs resonance, we directly display the results for the general case in  figures \ref{fig:lop-S2} and \ref{fig:ID-lop-S2}. Since the scalar, $S$, is now the lightest dark matter particle, one may expect some similarities with the singlet scalar model. Currently, this model is consistent only at the Higgs resonance and for  $M_S\gtrsim 950$ GeV.  From figure \ref{fig:lop-S2} we see that in the $Z_4$ model, instead, viable models span the whole range of  $M_S$ above the Higgs mass. The restriction $M_S>M_h$ is a consequence of the semiannihilation process $S+\psi\to \bar\psi+h$, which plays a complementary role in the determination of the $S$ relic density.

Even though  $\psi$ is the heavier dark matter particle, it  gives the dominant contribution to the dark matter density for $M_S\lesssim 900$ GeV -see the top-left panel of figure \ref{fig:lop-S2}. Above that mass, the $\psi$ fraction might decrease to just below $10\%$, and either dark matter particle could be dominant. From the top-right panel we conclude that the ratio of dark matter masses, $M_\psi/M_S$, can take any value (unlike for the regime $M_\psi<M_S$) and that the $S$ relic density is not  entirely driven by annihilations. Although  rarely dominant, semiannihilations are crucial to  open up the parameter space for  $M_S\lesssim 950$ GeV.

The center and bottom panels illustrate the viable ranges for the three couplings. It is clear that $y_p$ can take pretty much any value while $\lambda_{SH}$ and $y_s$ vary over a narrow band and tend to increase with $M_S$. In our scans, the maximum allowed values of $M_S$ and $M_\psi$ are $2$ and $6$ TeV respectively. From the figures, one can see that when $M_S\sim 2$ TeV, the couplings $\lambda_{SH}$ and $y_s$ have not converged to $3$ (the maximum allowed), indicating that it is possible to go to slightly larger masses. In any case, the  most interesting region is $M_S\lesssim 950$ GeV, where the singlet scalar model is excluded but our model is not.

The detection prospects are demonstrated in figure \ref{fig:ID-lop-S2}. The top panels show the scaterring cross section off nuclei for the fermion (left) and the scalar (right). In both cases, most points in our scan lie within the sensitivity of DARWIN. In fact, for the scalar there are practically no points that could escape future detection. Direct detection thus provides a reliable way to test this scenario. Regarding indirect detection (bottom panel), the most promising process is $\psi+\psi\to S+h$, with a gamma-ray flux arising from the decay of the Higgs.  From the figure we see that no points are currently excluded.

\section{Discussion}\label{sec:discussion}
We have seen in the previous sections that the $Z_4$ model  with a singlet fermion and a  singlet real scalar is a simple, predictive and testable scenario to explain the dark matter. Here we want to demonstrate that this framework can be straightforwardly generalized to other $Z_N$ symmetries and to additional dark matter particles.   

Under a $Z_N$ symmetry the operator $\overline{\psi^c}\psi S$ is invariant if the condition that the product of the field charges gives $1$ is fulfilled. In other words if  
\begin{align}
    &I:\,\, 2n_\psi+n_S=N,\,\,\,\text{or},\\
    &II:\,\,2n_\psi+n_S=0,
\end{align} 
where $Z_{N}(\psi)= (w_{N})^{n_\psi}$ and $Z_{N}(S)= (w_{N})^{n_S}$. Since for $n_S=N/2$ the scalar field remains as a real one,  the condition $I$ implies $n_\psi=N/4$ whereas $II$ demands $n_\psi=-N/4$ (that is $Z_N(\psi^c)=3N/4$). 

It follows that the $Z_4$ symmetry can be exchanged by a larger symmetry $Z_{4n}$ with the charges of the fermion and scalar dark matter particles given by 
\begin{align}
    Z_{4n}(\psi)= w_{4n}^{n}=i,\hspace{1cm}     Z_{4n}(S)= w_{4n}^{2n}=w_{4n}^{-2n}=-1,
\end{align}
where $w_{4n}^{4n}=1$. In this way the fermion and scalar fields remain as a Dirac fermion and real scalar, respectively, and  both conditions $I$ and $II$ are realized. The $Z_4$ model we studied is thus the lowest $Z_N$ model with a real scalar and a fermion, and the results of our analysis directly apply to other equivalent scenarios.    

For $Z_N$ symmetries with $N\neq 4n$ the real scalar field must be promoted to a complex field and, depending on the transformation properties, the interaction Lagrangian can take two possible structures
\begin{align}
 \mathcal{L}^{I[II]}&=\,\,\frac{1}{2}\left(y_s\overline{\psi^c}\psi + y_{p}\overline{\psi^c}\gamma_5\psi \right) S^{[*]}+ \rm{h.c.}.
 \end{align}
The simplest case is realized through a $Z_3$ symmetry since in that case both conditions $I$ and $II$ are equivalent. In follows that the fermion and scalar fields transform under the $Z_3$ symmetry in the same way, that is  
\begin{align}
 Z_{3}(\psi)=Z_{3}(S)= w_{3}. 
\end{align}
The interaction Lagrangian, however, comprises both possible structures $ \mathcal{L}_{Z_3}= \mathcal{L}^{I}+ \mathcal{L}^{II}$, leading to a larger set of free parameters. 
In the case $N=5$, the two possible charge assigments for $\psi$ and $S$ are   
\begin{align}
I: &\,\,\,    Z_{5}(\psi)= w_{5}^{2},\hspace{1cm}     Z_{5}(S)= w_{5}^1;\\ 
II:&\,\,\,    Z_{5}(\psi)= w_{5}^{1},\hspace{1cm}     Z_{5}(S)= w_{5}^{-2}. 
\end{align}
Similarly, for the case $N=6$, the fields must transform as 
\begin{align}
I: &\,\,\,    Z_{6}(\psi)= w_{6}^{2},\hspace{1cm}     Z_{6}(S)= w_{6}^2;\\ 
II:&\,\,\,    Z_{6}(\psi)= w_{6}^{1},\hspace{1cm}     Z_{6}(S)= w_{6}^{-2}. 
\end{align}
The case $N=7$ admits three possible charge assignments, but two of them are actually equivalent:
\begin{align}
IA: &\,\,\,    Z_{7}(\psi)= w_{7}^{2},\hspace{1cm}     Z_{7}(S)= w_{7}^3;\\ 
IB: &\,\,\,    Z_{7}(\psi)= w_{7}^{3},\hspace{1cm}     Z_{7}(S)= w_{7}^1;\\ 
II:&\,\,\,    Z_{7}(\psi)= w_{7}^{1},\hspace{1cm}     Z_{7}(S)= w_{7}^{-2}. 
\end{align}
In this way, the $Z_4$ symmetry of our model can be generalized to other $Z_N$'s. Notice that once the scalar field is promoted to be complex, larger values for the relevant couplings ($\lambda_{SH},y_s,y_p$) are required due to the extra degree of freedom contributing to the $S$ relic abundance.   On the other hand, the scenarios based on a $Z_{3N}$ symmetry where the scalar field has a charge $w_{3N}^N$ bring along an extra cubic interaction term  $S^3$, which leads to $S$ semi-annihilation processes that can significantly decrease the relic density of the scalar particle~\cite{Yaguna:2021vhb}.       

One can also  envision scenarios in which these $Z_N$ symmetries are actually remnants of a new $U(1)_X$ gauge symmetry. In this case $S$ must be a complex field and the condition $II$  becomes mandatory --in order to have the same interaction Lagrangian. In addition, the charges of $\psi$, $S$, and $\phi$ (the additional scalar  required to breakdown the $U(1)_X$ symmetry) must fulfill 
\begin{align}
    2q_\psi=-q_S,\hspace{1cm}|q_S|\neq |q_\phi|\neq 2|q_\phi|\neq 3|q_\phi|. 
\end{align}

Another way to extend these $Z_N$ models is via additional dark matter particles. With a $Z_4$ symmetry, for instance, it is possible to incorporate an extra complex scalar field, $S_2$, with charge $Z_4(S_2)=i$, leading to a three-component dark matter scenario. More interesting seems to be the scenario with two fermions and one complex scalar, which can be realized under a $Z_5$ (or higher) symmetry with charges $Z_5(\psi_1)=w_5^1$, $Z_5(\psi_2)=w_5^2$ and $Z_5(S)=w_5^2$. In such a scenario the interaction Lagrangian couples both fermion dark matter fields, $\overline{\psi_1^c}\psi_2 S$. Notice that a different charge assignment for the scalar $Z_5(S)=w_5^1$ leads to the interaction Lagrangian $\overline{\psi_1}\psi_2 S^*$.    
On the other hand, a $Z_6$ scenario allows us to have an interaction term for each fermion field via the charge assignment $Z_6(\psi_1)=w_6^1$, $Z_6(\psi_2)=w_6^2$ and $Z_6(S)=w_6^2$. 
By the same token, scenarios with even more dark matter particles could be obtained.

With respect to the $Z_N$ scenarios for \emph{scalar} dark matter \cite{Yaguna:2019cvp}, these new scenarios with both fermion and scalar dark matter feature two crucial advantages. First, they tend to be simpler as they typically introduce fewer free parameters --the fermion interactions are more restricted. Second, a smaller $Z_N$ symmetry can often be used. To obtain a two (three)-component dark matter scenario with only scalars requires a $Z_4(Z_6)$ symmetry, whereas with  fermions and scalars a $Z_3(Z_4)$ is enough, as shown above. 

This brief discussion makes clear  that the $Z_4$ model we investigated belongs to a large class of multi-component dark matter models in which the dark matter particles are fermions and scalars that are stabilized by a single $Z_N$ symmetry. From a different perspective, this class of models can be considered as ultraviolet realizations of the standard fermionic Higgs portals~\cite{Patt:2006fw} 
\begin{align}
    \mathcal{O}_1=(H^\dagger H)(\overline{\psi}\psi),\,\,\,    \mathcal{O}_2=(H^\dagger H)(\overline{\psi}\gamma_5\psi),
\end{align}
as well of the $d=5$ operators
\begin{align}
    \mathcal{O}_3=(H^\dagger H)(\overline{\psi^c}\psi),\,\,\,    \mathcal{O}_4=(H^\dagger H)(\overline{\psi^c}\gamma_5\psi).
\end{align}
The phenomenology of most of these models has yet to be studied in detail. 

\section{Conclusions}\label{sec:conclusions}
In this paper we reconsidered the scenario proposed in \cite{Cai:2015zza} --a two-component dark matter model in which the dark matter particles,  a singlet fermion ($\psi$) and a singlet scalar ($S$), are stabilized by a single $Z_4$ symmetry. The phenomenology of this model  is controlled by just five parameters: the two dark matter masses ($M_\psi, M_S$) and three dimensionless couplings ($\lambda_{SH}, y_s, y_p$). For the first time, we incorporated the pseudoscalar coupling ($y_p$) in the analysis, and found that it has a significant  impact on the viable regions --compare e.g.  figures \ref{fig:ypnull} and \ref{fig:yyy}. We investigated how these parameters affect dark matter observables  and obtained, for each regime ($M_\psi<M_S$ and $M_S<M_\psi$), a large sample of models consistent with all current bounds, including the most recent direct detection limits, which are quite relevant. Our analysis confirmed the essential role that semiannihilations play in obtaining the relic density, a point already stressed in  \cite{Cai:2015zza}, but also uncovered novel and important facts about this model, such as: $i)$  dark matter masses below $1$ TeV or so are allowed for both regimes; $ii)$ the fermion gives the dominant contribution to the relic density when $M_\psi<M_S$ and also when $M_S<M_\psi$ and $M_S<900$ GeV; $iii)$ the fermion direct detection cross section is detectable in spite of being generated at 1-loop; $iv)$ both dark matter particles could be observed in planned direct detection experiments, providing a  promising way to probe the model and to differentiate it from more conventional scenarios.  In addition, we characterized in detail the  allowed regions of this model, and studied the prospects for the direct and indirect detection of dark matter. Finally, we showed that  this model can straightforwardly be extended to other  $Z_N$ symmetries and to additional dark matter particles. In conclusion, we demonstrated that the $Z_4$ model with a singlet fermion and a real singlet scalar currently offers a predictive and well-motivated alternative to explain the dark matter.  The model not only is compatible with all present bounds but it also yields observable signals in ongoing and planned dark matter detectors.

\section*{Acknowledgments}
The work of OZ is supported by Sostenibilidad-UdeA and the UdeA/CODI Grants 2017-16286 and 2020-33177. 

\section*{Appendix}
In this section we report the expressions for terms associated to the cross sections of the relevant dark matter processes discussed in section~\ref{sec:DMpheno}. The cross section for the self annihilation process $\psi\psi\to Sh$ involves the parameters 
\begin{align}
\Delta&=256M_\psi^2(M_S^2-4M_\psi^2)^3\sqrt{M_h^4+(M_S^2-4M_\psi^2)^2-2M_h^2(M_S^2+4M_\psi^2)},\\
C_p &= (M_S^3-M_h^2M_S)^2 -4(3M_h^4-5M_h^2M_S^2+4M_S^4)M_\psi^2 + 80(M_h^2+M_S^2)M_\psi^4-128M_\psi^6,\\  
C_s&=(M_S^2-4M_\psi^2)((M_h-M_S)^2-4M_\psi^2)((M_h+M_S)^2-4M_\psi^2),
\end{align}
whereas in the cross section for semiannihilation process $\psi S\to \bar{\psi}h$  enter the parameters
\begin{align}
    \Delta' &= 192\pi M_S M_\psi [M_S^3+M_\psi(2M_S^2-M_h^2) ]^4 \sqrt{(M_S^2-M_h^2)((M_S+2M_\psi)^2-M_h^2)},\\
    C'_s & = 
   M_h^8 M_{\psi } \left(9 M_S M_{\psi }+4 M_S^2+3 M_{\psi }^2\right)\nonumber\\
   &-2 M_h^6
   \left(25 M_S^4 M_{\psi }+38 M_S^3 M_{\psi }^2+23 M_S^2 M_{\psi }^3+15 M_S M_{\psi
   }^4+5 M_S^5+6 M_{\psi }^5\right)\nonumber\\
   &+M_h^4 M_S^2 \left(211 M_S^4 M_{\psi }+359 M_S^3
   M_{\psi }^2+307 M_S^2 M_{\psi }^3+174 M_S M_{\psi }^4+45 M_S^5+48 M_{\psi }^5\right)\nonumber\\
   &-2M_h^2 M_S^4 \left(M_S+2 M_{\psi }\right) \left(3 M_S+2 M_{\psi }\right) \left(28 M_S^2
   M_{\psi }+33 M_S M_{\psi }^2+9 M_S^3+18 M_{\psi }^3\right)\nonumber\\
   &+M_S^6 \left(M_S+2 M_{\psi
   }\right){}^2 \left(59 M_S^2 M_{\psi }+66 M_S M_{\psi }^2+19 M_S^3+24 M_{\psi
   }^3\right),\\
    C'_p & = M_h^8 M_{\psi } \left(9 M_S
   M_{\psi }+4 M_S^2+3 M_{\psi }^2\right)\nonumber\\
   &-2 M_h^6 \left(25 M_S^4 M_{\psi }+40 M_S^3
   M_{\psi }^2+31 M_S^2 M_{\psi }^3+21 M_S M_{\psi }^4+5 M_S^5+6 M_{\psi }^5\right)\nonumber\\
   &+M_h^4
   M_S^2 \left(203 M_S^4 M_{\psi }+379 M_S^3 M_{\psi }^2+411 M_S^2 M_{\psi }^3+266 M_S
   M_{\psi }^4+45 M_S^5+64 M_{\psi }^5\right)\nonumber\\
   &-2 M_h^2 M_S^4 \left(M_S+2 M_{\psi }\right)
   \left(94 M_S^3 M_{\psi }+169 M_S^2 M_{\psi }^2+136 M_S M_{\psi }^3+21 M_S^4+36 M_{\psi
   }^4\right)\nonumber\\
   &+M_S^6 \left(M_S+2 M_{\psi }\right){}^3 \left(13 M_S M_{\psi }+7 M_S^2+4
   M_{\psi }^2\right).
\end{align}
Finally, the diferential cross section for the dark matter conversion process $\bar{\psi}\psi \to SS$ depends on  
\begin{align}
    \Sigma_t=&-y_p^4\left[\left(t-M_{\psi }^2\right) \left(-M_{\psi }^2+s+t\right)+2 M_S^2
   \left(M_{\psi }^2-t\right)+M_S^4\right]\nonumber\\
   &-2 y_p^2 y_s^2 \left[\left(M_{\psi }^2+3
   t\right) \left(-M_{\psi }^2+s+t\right)-6 M_S^2 \left(M_{\psi }^2+t\right)+3
   M_S^4\right]\nonumber\\
   &-y_s^4 \left[-s M_{\psi }^2+\left(M_S^2-3 M_{\psi }^2\right) \left(M_S^2-3
   M_{\psi }^2-2 t\right)+t (s+t)\right],\\
  \Sigma_u=& -y_p^4\left[\left(t-M_{\psi }^2\right) \left(-M_{\psi }^2+s+t\right)+2 M_S^2
   \left(M_{\psi }^2-t\right)+M_S^4\right]\nonumber\\
   &-2 y_p^2 y_s^2 \left[\left(t-M_{\psi
   }^2\right) \left(3 (s+t)-7 M_{\psi }^2\right)+2 M_S^2 \left(M_{\psi }^2-3 t\right)+3
   M_S^4\right]\nonumber\\
   &-y_s^4 \left[-9 s M_{\psi }^2-\left(M_S^2+5 M_{\psi }^2\right)
   \left(-M_S^2-5 M_{\psi }^2+2 t\right)+t (s+t)\right],\\
   \Sigma_{tu}=&-y_p^4\left[\left(t-M_{\psi }^2\right) \left(-M_{\psi }^2+s+t\right)+2 M_S^2
   \left(M_{\psi }^2-t\right)+M_S^4\right]\nonumber\\
   &+2 y_p^2 y_s^2 \left[M_{\psi }^2 \left(2
   M_S^2+s+6 t\right)-3 \left(-2 t M_S^2+M_S^4+t (s+t)\right)-3 M_{\psi }^4\right]\nonumber\\
   &-y_s^4
   \left[\left(3 M_{\psi }^2+t\right) \left(-5 M_{\psi }^2+s+t\right)+2 M_S^2
   \left(M_{\psi }^2-t\right)+M_S^4\right].
\end{align}
\bibliography{references}

\end{document}